%% file: nv.tex
\documentclass[10pt,twocolumn]{article}
\usepackage[affil-it]{authblk}
\usepackage{geometry}               
\usepackage{graphicx}
\usepackage{amssymb}
\usepackage{amsmath}
\usepackage{epstopdf}
\usepackage[cm]{fullpage}
\usepackage{cite}
\usepackage{bm}
\usepackage{cancel}
\usepackage{array}

\title{Thermal quantum time-correlation functions from classical-like dynamics}
\author{Timothy J.\ H.\ Hele\thanks{On intermission from: Jesus College, University of Cambridge, UK.}}
\affil{Department of Chemistry and Chemical Biology, Cornell University, Ithaca, New York 14853, USA}

\input{abbrev}
\begin{document}
\newcommand{\cev}[1]{\reflectbox{\ensuremath{\vec{\reflectbox{\ensuremath{#1}}}}}}
\newcommand{\Lc}{\mL_{\mathrm{C}}}
\newcommand{\LM}{\mL^{[M]}}

\bibliographystyle{tim}
\maketitle

\begin{abstract}
Thermal quantum time-correlation functions are of fundamental importance in quantum dynamics, allowing experimentally-measurable properties such as reaction rates, diffusion constants and vibrational spectra to be computed from first principles. Since the exact quantum solution scales exponentially with system size, there has been considerable effort in formulating reliable linear-scaling methods involving exact quantum statistics and approximate quantum dynamics modelled with classical-like trajectories. Here we review recent progress in the field with the development of methods including Centroid Molecular Dynamics (CMD), Ring Polymer Molecular Dynamics (RPMD) and Thermostatted RPMD (TRPMD). We show how these methods have recently been obtained from `Matsubara dynamics', a form of semiclassical dynamics which conserves the quantum Boltzmann distribution. We also rederive \shortt\ quantum transition-state theory (QTST) in the Matsubara dynamics formalism showing that Matsubara-TST, like RPMD-TST, is equivalent to QTST. We end by surveying areas for future progress. \emph{Submitted as a New View article to Molecular Physics \emph{(www.tandfonline.com/toc/tmph20/current)} on 11th January 2017.}
\end{abstract}

\section{Introduction}
Quantum thermal time-correlation functions \cite{zwa01a,nit06a} are routinely used to calculate reaction rates, spectra and diffusion constants amongst many other physically observable quantities, and provide a useful bridge between the algebra of quantum mechanics and experimental measurement. In general they can only be computed exactly for very small or model systems, and there is consequently a need for reliable approximate computation with classical-like scaling (i.e.\ linear scaling w.r.t.\ the number of dimensions of the system). The purpose of this New View article is to review the origins of a number of these methods; namely the approximations they make to the exact quantum evolution and the conditions under which they are likely to be valid. This should allow a theoretician to discern for themselves the optimal method for a given problem. 

This article is designed to provide an overview of the field with references for further reading and is not intended to be exhaustive. Applications of many of the methods discussed here have already been extensively reviewed, including centroid molecular dynamics (CMD) \cite{vot96a}, ring polymer molecular dynamics (RPMD) \cite{hab13a}, RPMD rate theory \cite{sul16a} and the linearized semiclassical initial-value representation (LSC-IVR) \cite{liu15a}. Consequently, applications of these methods are only mentioned when pertinent. 

We also rederive quantum transition-state theory (QTST) in the Matsubara formalism, showing that Matsubara TST is identical to QTST provided that the dividing surface is only a function of the Matsubara modes, and which in turn is identical to RPMD-TST when the dividing surface is invariant to cyclic permutation in imaginary time. For reviews on rate theory more generally, see Refs.~\cite{tru96a,pol05a,nym14a,han90a}.

There exist many other methods to simulate quantum dynamics which are not covered here, including exact quantum methods such as multi-configuration time-dependent Hartree (MCTDH) \cite{bec00a}, matrix-based methods \cite{bow08a}, and path-integrals \cite{mak95a}. Other approaches include gaussian wavepacket propagation \cite{sha01a}, semiclassical dynamics \cite{mil01a, tho04a} and mixed quantum-classical dynamics \cite{ant15a,kap06a,kap99a}.

For most of the article we assume that dynamics is on a single Born-Oppenheimer potential energy surface that is known and differentiable (either of a model form, fitted to some set of parameters, or from ab initio electronic structure theory); the computation of accurate potential energy surfaces is a discipline in itself. We touch upon extensions to non-adiabatic dynamics towards the end. We generally assume that the systems being described are in thermal equilibrium; application to non-equilibrium systems is an interesting area of present research \cite{wel16a}.

The article is structured as follows. In section~\ref{sec:tcf} we review classical and quantum thermal time-correlation functions, the Wigner transform and the Moyal series. Section~\ref{sec:lscivr} touches upon LSC-IVR, and section~\ref{sec:mat} provides the derivation of Matsubara dynamics. Section~\ref{sec:approx} covers approximations to Matsubara dynamics such as CMD, RPMD and TRPMD, and section~\ref{sec:qtst} gives an alternative derivation of QTST in the Moyal/Matsubara formalism. Section~\ref{sec:fut} presents directions for future research and section~\ref{sec:con} concludes.

\section{Thermal time-correlation functions}
\label{sec:tcf}
Here we briefly present background theory sufficient to follow the remainder of the article; further detail is available in standard texts\cite{fre02a, nit06a, zwa01a}.
\subsection{Classical}
For simplicity we consider a one-dimensional system, extension to further dimensions being straightforward\cite{nit06a}, with position $q$ and momentum $p$ and a classical Hamiltonian
\begin{align}
H(p,q) = \frac{p^2}{2m} + V(q).
\end{align}
The thermal correlation function between observables $A$ and $B$ at inverse temperature $\beta \equiv 1/k_{\rm B}T$ (where $k_{\rm B}$ is the Boltzmann constant) is generally written as
\begin{align}
G_{AB}(t) = \tph \int dp \int dq \ e^{-\beta H(p,q)} A(p,q) B(p_t,q_t) \eql{conv}
\end{align}
where $p$ and $q$ are sampled at zero time and $(p_t,q_t) \equiv (p_t(p,q,t),q_t(p,q,t))$ are the solutions to a classical trajectory for length $t$ starting at $(p,q)$ at time $t=0$. The correlation function can equivalently be given as
\begin{align}
G_{AB}(t) = \tph \int dp \int dq \ e^{-\beta H(p,q)} A(p,q) B(p,q,t) \eql{equiv}
\end{align}
where $B(p,q,t)$ corresponds to an initial phase-space distribution $(p,q)$ propagated for time $t$. Formally, one can obtain the dynamical equations of motion by differentiating \eqr{equiv} w.r.t.\ time to obtain
\begin{align}
\dd{}{t} B(p,q,t) = & \ddp{B(p,q,t)}{q} \dd{q}{t} + \ddp{B(p,q,t)}{p} \dd{p}{t} \\
= & \ddp{B(p,q,t)}{q} \frac{p}{m} - \ddp{B(p,q,t)}{p} \ddp{V(q)}{q} \eql{timev}
\end{align}
where we have applied Newton's first and second law to obtain \eqr{timev}. Strictly speaking, we are also assuming that the observables themselves are not explicit functions of time, i.e.
\begin{align}
\left(\ddp{B(p,q,t)}{t}\right)_{p,q} = 0,
\end{align}
and likewise for $A$, which is the case for all correlation functions considered in this article. 
Equation~\eqref{eq:timev} allows us to define a classical \emph{Liouvillian}\footnote{Following the convention of Zwanzig \cite{zwa01a} we define the Liouvillian without a prefactor of $i$.}
\begin{align}
\mL = \frac{p}{m} \ddp{}{q} - \ddp{V(q)}{q} \ddp{}{p} \eql{lclas}
\end{align}
such that 
\begin{align}
\dd{}{t} B(p,q,t) = \mL B(p,q,t) \eql{tl}
\end{align}
which has a formal solution $B(p,q,t) = e^{\mL t} B(p,q,0)$ and therefore
\begin{align}
G_{AB}(t) = \tph \int dp \int dq \ e^{-\beta H(p,q)} A(p,q) e^{\mL t} B(p,q,0).
\end{align}
To see how \eqr{conv} is equivalent to \eqr{equiv} we differentiate \eqr{conv} w.r.t.\ $t$, obtaining
\begin{align}
\dd{}{t} B(p_t,q_t) = \ddp{B(p_t,q_t)}{q_t} \dd{q_t}{t} + \ddp{B(p_t,q_t)}{p_t} \dd{p_t}{t} \eql{dtb}
\end{align}
but if \eqr{conv} is a solution to \eqr{equiv} then by \eqr{tl}, the LHS of \eqr{dtb} must be equal to the action of the Liouvillian on $B(p_t,q_t)$, which is
\begin{align}
\mL B(p_t,q_t) = \ddp{B(p_t,q_t)}{q_t} \mL q_t + \ddp{B(p_t,q_t)}{p_t}\mL p_t. \eql{lb}
\end{align}
Comparing \eqr{dtb} and \eqr{lb} gives 
\begin{align}
\dd{q_t}{t} = \mL q_t, \qquad \dd{p_t}{t} = \mL p_t
\end{align}
which have formal solutions $q_t = e^{\mL t} q, \ p_t = e^{\mL t} p$. This means that instead of propagating a phase space density in $B(p,q,t)$, one can simply propagate individual positions and momenta to find $(p_t,q_t)$ and insert into the function $B(p_t,q_t)$, which is computationally easier. However, if $\mL$ contains higher derivatives in $p$ and/or $q$ (as is the case in exact quantum evolution and stochastic dynamics) then this convenient property no longer holds.

If $B = H$, then from \eqr{lclas} $\mL H = 0$, meaning that classical dynamics conserves the classical Hamiltonian, as to be expected. It follows that $\mL e^{-\beta H(p,q)}=0$ and the classical dynamics conserves the classical Boltzmann distribution. 

If we differentiate \eqr{equiv} w.r.t.\ $t$, apply \eqr{lclas} use integration by parts on the derivatives in $p$ and $q$ we obtain
\begin{align}
\dd{}{t} G_{AB}(t) = -\tph \int dp \int dq \ e^{-\beta H(p,q)} A(p,q) \ola \mL B(p,q,t) \eql {intpa}
\end{align}
where $\ola \mL$ is `acting backwards' onto $e^{-\beta H(p,q)} A(p,q)$, but using the product rule and that $\mL e^{-\beta H(p,q)} = 0$, this gives
\begin{align}
\dd{}{t} G_{AB}(t) = -\tph \int dp \int dq \ e^{-\beta H(p,q)} B(p,q,t) \mL A(p,q).
\end{align}
Integration of this, noting that $B(p,q,0)=B(p,q)$ gives 
\begin{align}
G_{AB}(t) = & \tph \int dp \int dq \ e^{-\beta H(p,q)} B(p,q) e^{-\mL t} A(p,q) \no \\
= & G_{BA}(-t)
\end{align}
which is \emph{detailed balance}. Note that this is a stronger condition than time reversal symmetry, which only implies [from \eqr{intpa}] that
\begin{align}
G_{AB}(t) = & \tph \int dp \int dq \ B(p,q) e^{-\mL t} [e^{-\beta H(p,q)} A(p,q)]
\end{align}
where the distribution has to be propagated too. In general, if the dynamics conserves the distribution then the correlation function will observe detailed balance (strictly speaking, for stochastic systems this is a necessary but not sufficient requirement \cite{gar09a,hel15c}).

\subsection{Quantum}
Similar to the classical case, we consider a one-dimensional system with mass $m$, co-ordinate $q$ with conjugate momentum $p$ and quantum Hamiltonian
\begin{align}
 \hat H = \frac{\hat p^2}{2m} + V(\hat q). \eql{qmham}
\end{align}
In this section we introduce a variety of quantum time-correlation functions and briefly discuss their properties, particularly concerning the ease with which they may be approximated by classical-like dynamics. 

\subsubsection{Conventional time-correlation function}
The conventional quantum time-correlation function is given by\cite{nit06a, tuc10a} 
\begin{align}
 c_{AB}(t) = \tr \left[\eb \hat A \etb \hat B \etf\right] \eql{cabt}
\end{align}
such that $c_{AB}(0) = \tr [\eb \hat A\hat B]$, giving the thermal average of $\hat A $ and $\hat B$. Since $[\etf,\hat H]=0$, $c_{AH}(t) = c_{AH}(0)$ and the quantum dynamics conserves the quantum Hamiltonian. 

This is sometimes called the `asymmetric-split' correlation function, since the Boltzmann operator is placed asymmetrically on one side of $\hat A$. To picture this function as in \figr{corfun}a we insert identities into \eqr{cabt}, which when $\hat A$ and $\hat B$ are functions of position only gives \cite{tuc10a}
\begin{align}
c_{AB}(t) = & \int dx \int dy \int dz \bra{x} \eb \ket{y} A(y) \no \\
& \times \bra{y} \etb \ket{z} B(z) \bra{z} \etf \ket{x} \eql{idin}
\end{align}
We can therefore imagine starting from point $x$ in \figr{corfun}a and taking an imaginary time path $\eb$ ending at $y$, at which $A(y)$ is evaluated. We then take a backwards real time path $\etb$ from $y$ to $z$, at which $B(z)$ is evaluated, followed by a real time path $\etf$ from $z$ to $x$, completing the trace.

However, the correlation function is not necessarily real, even for an autocorrelation function (where $\hat A = \hat B$); one can show by exploiting $[\eb, e^{\pm i \hat H t/\hbar}]=0$ that for arbitrary $\hat A$ and $\hat B$
\begin{align}
 c_{AB}(-t)^* = c_{BA}(t).
\end{align}

\begin{figure}[tb]
\centering
\includegraphics[width=\columnwidth]{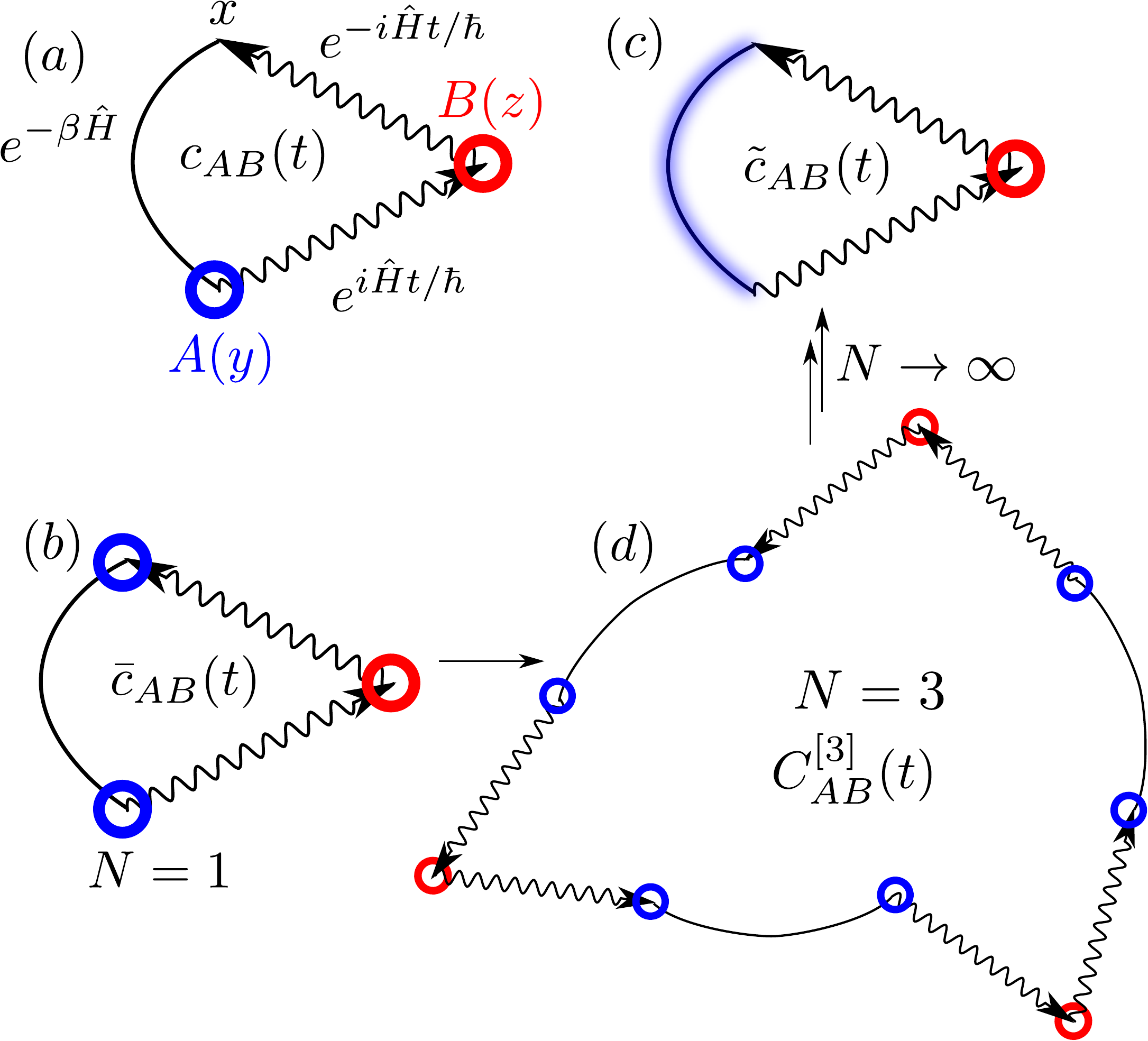}
\caption{Different forms of quantum mechanical correlation functions discussed in this article. The imaginary time path shown as a curved line, the real time path with a wavy line and the $\hat A$ and $\hat B$ operators as blue and red circles respectively. (a) is the conventional asymmetric-split correlation function, (b) one form of symmetric splitting, (c) the Kubo-transformed function (with the operator $\hat A$ `smeared' along the imaginary time trajectory). The Generalized Kubo transformed function (d) is obtained by polymerising (b), and is equivalent to the conventional Kubo transformed function (c) in the $N\to \infty$ limit for linear operators. }
\figl{corfun}
\end{figure}

\subsubsection{Symmetric-split time-correlation function}
Since \eqr{cabt} can be complex and the classical correlation function is not, we wish to rewrite \eqr{cabt} to be real. A simple way to do this would be to take the real part of \eqr{cabt}, giving
\begin{align}
\bar c_{AB}(t) =: & \Re  c_{AB}(t) \no \\
= & \tr \left[\frac{1}{2}(\hat A \eb + \eb \hat A) \etb \hat B \etf\right] \eql{symcab}
\end{align}
which is pictured in \figr{corfun}b. Although this looks more complex that \eqr{cabt}, if we insert identities as in \eqr{idin} and then change to sum-and-difference variables $q = (x+y)/2$, $\Delta = y-x$, noting that the Jacobian of the transformation is unity, we obtain (for $\hat A$ which is a linear function of $\hat x$)
\begin{align}
\bar c_{AB}(t) = & \int dq \int d\Delta \int dz \bra{q-\Delta/2}\eb \ket{q+\Delta/2} A(q) \no \\
 & \times \bra{q+\Delta/2} \etb \ket{z} B(z) \bra{z} \etf \ket{q - \Delta/2}. \eql{symcab2}
\end{align}
We can, for linear operators, consider $\hat A$ to be acting at the mid-point of the imaginary time trajectory (this can also hold for some nonlinear operators, see Section~\ref{sec:qtst}).

\subsubsection{Kubo-transformed time-correlation function}
Although \eqr{symcab} is real and therefore an improvment upon \eqr{cabt} for classical approximation, the action of $\hat A$ at specific points in imaginary time (rather than smoothed over all points) leads to difficulties with classical approximations, as we shall see later. A correlation function which treats all points in imaginary time equally is the Kubo-transformed correlation function \cite{kub57a}
\begin{align}
 \ti c_{AB}(t) = \frac{1}{\beta}\int_0^\beta d\lambda \ \tr[e^{-(\beta - \lambda) \hat H} \hat A e^{-\lambda \hat H} \etb \hat B \etf] \eql{cabkub}
\end{align}
which corresponds to the zero-time operator $\hat A$ being `smeared' through the imaginary time operator $\eb$, as pictured in \figr{corfun}c. This can be obtained for some quantum mechanical properties using linear response theory \cite{yam60a}. In addition to the symmetry properties for \eqr{cabt}, by switching integration limits one can show that the Kubo transformed correlation function is always real,
\begin{align}
 \ti c_{AB}(t) = \ti c_{AB}(t)^* 
\end{align}
and that it obeys detailed balance, i.e.
\begin{align}
 \ti c_{AB}(-t) = \ti c_{BA}(t)
\end{align}
and so is more `classical' than the correlation function in \eqr{cabt}. Further symmetry properties of these correlation functions are given in Ref.~\cite{cra04a}.

\subsubsection{Generalized Kubo-transformed time-correlation function}
It is possible to rewrite the Kubo-transformed correlation function in a more symmetric form, known as the Generalized Kubo Transformed correlation function \cite{hel13a,alt13a,hel13b,hel14a}. To sketch how this comes about, consider dividing up the imaginary time trajectory $\eb$ in the symmetric-split \eqr{symcab2} into $N$ chunks, and at each chunk inserting $\etf \etb$, as pictured in \figr{corfun}d for $N=3$. This gives 
\begin{align}
 &\CABNt = \int d\bq \int d\bDelta \no\\
 & \times \piNz \bra{q_{i-1} - \Delta_{i-1}/2}\frac{1}{2}(\hat A \ebN + \ebN \hat A) \ket{q_i + \Delta_i/2}\no \\
 & \qquad\times \bra{q_i + \Delta_i/2} \etb \hat B \etf \ket{q_i - \Delta_i/2} \eql{genk}
\end{align}
where (for linear $\hat A$ and $\hat B$)  
\begin{align}
 \hat A = \frac{1}{N}\smkNz \hat A_k \eql{lin}
\end{align}
with $\hat A_k$ acting on the $k$th path-integral `bead' and likewise for $\hat B$, where we loosely define $q_i$ to be the $i$th bead (see appendix~\ref{ap:rp} for a discussion of ring polymers and bead terminology). One can show (by evaluating the summations in the correlation function term-by-term and removing $\etb\etf = \hat 1$ identities) that with $\hat A$ and $\hat B$ defined as in \eqr{lin} then this is equal to the conventional Kubo transformed correlation function in the large $N$ limit \cite{hel15a}
\begin{align}
 \lNti \CABNt = \ti c_{AB}(t).
\end{align}
Nonlinear operators [which cannot easily be written as a sum like \eqr{lin}] are required for Quantum Transition-State Theory, and are detailed in Section~\ref{sec:qtst}. As we shall see later, the advantage of rewriting \eqr{cabkub} as the Generalized Kubo form is that the latter is symmetric with respect to permutation in imaginary time $\tau = \betaN\hbar$, corresponding to permuting the co-ordinates $q_i \to q_{i+1}$ \cite{hel15a}.

The above is not an exhaustive list of quantum time-correlation functions; there are theoretically infinitely may ways to split the zero-time operator within the Boltzmann distribution\cite{mil83a,hel13b}, one other common technique being $e^{-\beta \hat H/2} \hat A e^{-\beta \hat H/2}$ \cite{mil83a,tuc10a}.

By inserting energy eigenstates into \eqr{cabt} and \eqr{cabkub} one can relate the spectrum of the conventional and Kubo-transformed correlation functions\cite{cra04a,wit09a}
\begin{align}
 \ti I_{AB}(\omega) = \frac{1 - e^{-\beta\hbar\omega}}{\beta\hbar\omega} I_{AB}(\omega) 
\end{align}
where the spectrum is given by
\begin{align}
 \ti I_{AB}(\omega) = \frac{1}{2\pi} \inti dt \ e^{-i\omega t} \ti c_{AB}(t). \eql{ft}
\end{align}
and likewise for $I_{AB}(\omega)$.

\subsubsection{Applications}
To illustrate the scope of correlation functions, we now sketch how they may be used to compute diffusion, rates and spectra.

The diffusion constant is obtained as the integral of the Kubo-transformed velocity-velocity autocorrelation function\cite{mil05a}
\begin{align}
 D = \frac{1}{3Z} \int_0^\infty dt \ \ti c_{\mathbf{v} \cdot \mathbf{v}}(t)
\end{align}
where $Z$ is the partition function of the system. The rate constant can be obtained from the long-time limit of the flux-side time-correlation function\cite{mil74a,mil83a,yam60a,cha87a} (of the asymmetric, symmetric Kubo-transformed, and many other forms \cite{mil83a})
\begin{align}
 k_{\rm Q}(\beta) = \ltti \frac{1}{\Qrb}\ti c_{\rm fs}(t). \eql{kqb}
\end{align}
where $\Qrb$ is the partition function in the reactant region and the flux-side correlation function is
\begin{align}
 c_{\rm fs}(t) = \tr \left[\eb \hat F \etb h(\hat q - \qdd) \etf \right] \eql{cfsq}
\end{align}
although \eqr{kqb} also holds for the Kubo-transformed correlation function amongst others \cite{mil83a}. The flux operator is $\hat F = [\delta(\hat q - \qdd) \hat p + \hat p \delta(\hat q - \qdd)]/2m$ where $\delta(x)$ is the Dirac delta function and $\qdd$ is the location of the position-space dividing surface. Using the quantum mechanical continuity equation one can show that the exact quantum rate is independent of the location of the dividing surface \cite{cra05b}. The heaviside function $h(\hat q - \qdd)$ is defined such that
\begin{align}
h(q - \qdd) = 
\left\{
\begin{array}{cl}
1 & q \ge \qdd \\
0 & q < \qdd.
\end{array}
\right.
\end{align}
Since the flux operator is the time-derivative of the heaviside operator, the flux-side function is the integral of the flux-flux function \cite{mil83a}
\begin{align}
c_{\rm fs}(t) = \int_0^t c_{\rm ff}(t') dt' \eql{cff}
\end{align}
where $c_{\rm ff}(t)$ is obtained by changing $h(\hat q-\qdd)$ for $\hat F$ in \eqr{cfsq}, and $\cfs$ is minus the derivative of the side-side function
\begin{align}
c_{\rm fs}(t) = -\dd{}{t} c_{\rm ss}(t) \eql{css}
\end{align}
where $c_{\rm ss}(t)$ is obtained by changing $\hat F$ for $h(\hat q-\qdd)$ in \eqr{cfsq}. 
These identities, which generally hold for most classical flux-side time correlation functions too, will prove useful later. 

For infra-red spectra, the absorption coefficient is given as\cite{wit09a}
\begin{align}
 \alpha(\omega) = \frac{4\beta\pi^2\omega^2}{3\mathcal{V}c n(\omega)Z} \ti I_{\bm{\mu\mu}}(\omega) \eql{irabs}
\end{align}
where $\ti I_{\bm{\mu\mu}}(\omega)$ is the Kubo-transformed dipole autocorrelation function found using \eqr{ft}, $\mathcal{V}$ corresponds to the volume, $c$ the speed of light and $n(\omega)$ the refraction coefficient (approximately unity in the gas phase). 

The above is not exhaustive; other observables can be obtained from thermal quantum time-correlation functions such as neutron scattering \cite{cra06a}.

\subsection{Moyal series}
\label{sec:moy}
Having given the exact quantum time-correlation functions in the conventional operator representation, we now consider how the Wigner transform and Moyal series which can be used to rewrite correlation function in terms of phase-space positions and momenta. We use the conventional Kubo-transformed function in this section, but the derivation is equally applicable to the asymmetric or symmetric-split forms. 

Inserting position-space identities followed by changing to sum and difference variables as in \eqr{symcab2} gives 
\begin{align}
 c_{AB}(t) = & \int dq \int d\Delta \ \bra{q-\Delta/2} K_{\beta}(\hat A) \ket{q+\Delta/2} \no \\
 & \times \bra{q+\Delta/2} \hat B(t) \ket{q - \Delta/2}. \eql{qdtrans}
\end{align}
where we have abbreviated the Kubo transform as
\begin{align}
 K_{\beta}(\hat A) = \frac{1}{\beta} \int_0^\beta d\lambda \ e^{-(\beta - \lambda) \hat H} \hat A e^{-\lambda \hat H}
\end{align}
and $\hat B(t) = \etb \hat B \etf$ is the Heisenberg time-evolved $\hat B$. 
We can now insert another identity
\begin{align}
 1 = & \int d\Delta' \delta(\Delta + \Delta') \no \\
 = & \frac{1}{2\pi\hbar}\int d\Delta' \int dp\ e^{ip(\Delta + \Delta')/\hbar} 
\end{align}
where we have written the Dirac delta function on the first line as its Fourier transform on the second, and convert the $\Delta$ to $-\Delta'$ in the second bra-ket of \eqr{qdtrans}, giving
\begin{align}
 c_{AB}(t) = & \tph \int dq \int dp \no \\
 & \times \int d\Delta e^{ip\Delta/\hbar} \bra{q-\Delta/2} K_{\beta}(\hat A) \ket{q+\Delta/2} \no \\
 & \times \int d\Delta' e^{ip\Delta'/\hbar} \bra{q-\Delta'/2} \etb \hat B\etf \ket{q + \Delta'/2} \no \\
= &  \tph \int dq \int dp \ [K_{\beta}(\hat A)]_\rW (q,p) [B(t)]_\rW(q,p) \eql{cwig}
\end{align}
where $[\hat O ]_{\rW}$ defines the \emph{Wigner transform} of operator $\hat O$ \cite{wig32a}
\begin{align}
 [\hat O]_\rW (q,p) = \int d\Delta e^{ip\Delta/\hbar} \bra{q-\Delta/2} \hat O \ket{q+\Delta/2} \eql{wigtr}
\end{align}
All we have done in \eqr{qdtrans}--\eqr{wigtr} is to rewrite the correlation function is terms of classical-like phase-space variables $p$ and $q$. No approximation has been made, an in general solving \eqr{cwig} exactly is just as difficult as solving the original \eqr{cabkub}. The advantage of writing in a classical-like form is the ability to make approximations to the correlation functions such that they can be evaluated using classical or classical-like dynamics.

We now obtain the Liouvillian for a Wigner-transformed correlation function, starting by differentiating \eqr{cwig} w.r.t.\ $t$,
\begin{align}
 \dd{}{t} \ti c_{AB}(t) = \int dq \int dp \ [K_{\beta}(\hat A)]_\rW(q,p) \left[\frac{i}{\hbar}[\hat H, \hat B(t)] \right]_\rW(q,p)
\end{align}
where the commutator arises from noticing $\dd{}{t} \etb \hat B\etf = (i/\hbar)[\hat H, \etb \hat B \etf]$. The evaluation of the Wigner transform of the commutator is detailed in Ref.~\cite{hil84a} and here we give the main steps.

Using \eqr{qmham} we can write (dropping the prime on $\Delta'$ for simplicity)
\begin{subequations}
\begin{align}
& \dd{}{t} [B(t)]_\rW(q,p) = \no \\
& \frac{i}{\hbar} \int dp \ e^{ip\Delta/\hbar} \bra{q-\Delta/2}\left[\frac{\hat p^2}{2m},\hat B(t)\right] \ket{q+\Delta/2} \eql{kin} \\
 & + \frac{i}{\hbar} \int dp \ e^{ip\Delta/\hbar} \bra{q-\Delta/2}\left[V(q),\hat B(t)\right] \ket{q+\Delta/2}. \eql{pot}
\end{align}
\end{subequations}
Using the definition $\hat p = -i\hbar \dd{}{\hat q}$, we can take the position derivatives outisde the bra-kets, and using partial differentation show
\begin{align}
 \ddp{^2}{(q-\Delta/2)^2} - \ddp{^2}{(q+\Delta/2)^2} = -2 \ddp{}{q}\ddp{}{\Delta}
\end{align}
and using integration by parts $\dd{}{\Delta}$ can be converted into $ip/\hbar$. Combining the above into \eqr{kin} gives 
\begin{align}
 \frac{i}{\hbar} & \int dp \ e^{ip\Delta/\hbar} \bra{q-\Delta/2}\left[\frac{\hat p^2}{2m},\hat B(t)\right] \ket{q+\Delta/2} \no \\
 & = \frac{p}{m}\ddp{}{q} [\hat B(t)]_\rW \eql{kinterm}
\end{align}
which is Newton's first law. For the potential term in \eqr{pot}, we observe
\begin{align}
 V(q-\Delta/2) - V(q+\Delta/2) = -2\sinh\!\left(\frac{\Delta}{2} \ddp{}{q}\right).
\end{align}
Combining this with $\Delta$ being equivalent to $-i\hbar \dd{}{p}$ acting on the entire Wigner Transform we obtain
\begin{align}
 \frac{i}{\hbar} & \int dp \ e^{ip\Delta/\hbar} \bra{q-\Delta/2}\left[V(q),\hat B(t)\right] \ket{q+\Delta/2} \no\\
 & = -\frac{2}{\hbar} V(q) \sin\left( \frac{\hbar}{2} \frac{\ola \partial}{\partial q} \frac{\ora \partial}{\partial p} \right) [\hat B(t)]_\rW \eql{potterm}
\end{align}
where the arrows indicate in which direction the derivative acts, and which is like Newton's second law with higher-order terms in $\hbar$, as can be seen from expanding the sine series. 
Combining \eqr{kinterm} and \eqr{potterm} we obtain 
\begin{align}
 \dd{}{t} [\hat B(t)]_\rW = \LMoy [\hat B(t)]_\rW 
\end{align}
where $\LMoy$ is the \emph{Moyal series}\cite{gro46a,moy49a,hil84a}
\begin{align}
 \LMoy = \frac{p}{m}\ddp{}{q} - \frac{2}{\hbar} V(q) \sin\left( \frac{\hbar}{2} \frac{\ola \partial}{\partial q} \frac{\ora \partial}{\partial p} \right), \eql{lmoy}
\end{align}
which is referred to as a series since expanding the sine term gives a series in powers of $\hbar^2$. The correlation function is therefore
\begin{align}
 \ti c_{AB}(t) = \tph \int dq \int dp \ [K_{\beta}(\hat A)]_\rW (q,p) e^{\LMoy t} [B(0)]_\rW(q,p) \eql{cmoy}.
\end{align}
In general, computing the action of the Moyal series upon an obserable is as difficult as solving the Schr\"odinger equation by conventional matrix-based methods, due to the presence of the higher-order derivatives in \eqr{lmoy}, although there have been some approaches to address this \cite{kry09a}. In the following sections we therefore explore approximating the Moyal series or generalization of it to obtain classical-like dynamics.

\section{LSC-IVR}
\label{sec:lscivr}
Arguably the simplest way to approximate $\LMoy$ is to truncate in powers of $\hbar$, giving 
\begin{align}
 \mathcal{L}_0 = \frac{p}{m}\ddp{}{q} - \ddp{V(q)}{q}\ddp{}{p} \eql{l0}
\end{align}
which corresponds to purely classical evolution of the phase-space density from an initial quantum Boltzmann distribution, and has the appealing feature that the error from exact quantum evolution $\mL_{\rm Q}$ is known,
\begin{align}
\mL_{\rm Q} = & \LMoy - \mL_0 \no \\
= & \sum_{\nu=3{\rm,\ odd}} \left(\frac{i\hbar}{2}\right)^{\nu -1}\frac{1}{\nu!} \left( \frac{\ola \partial}{\partial q} \frac{\ora \partial}{\partial p} \right)^{\nu} \eql{lqdef}
\end{align}
which (by construction) only contains terms of $\mathcal{O}(\hbar^2)$ and higher. 
Inserting \eqr{l0} into the correlation function gives
\begin{align}
 \ti c_{AB}(t) \simeq & \int dp\int dq \ [K_{\beta}(\hat A)]_\rW (q,p) e^{\mL_0 t} [B(t)]_\rW(q,p) \no \\
 \equiv & \int dp\int dq \ [K_{\beta}(\hat A)]_\rW (q,p) [\hat B(0)]_\rW(q_t,p_t) \eql{lscivr}
\end{align}
where we have noted that, since $\mL_0$ is classical, it corresponds to inserting the time-evolved positions and momenta into $[\hat B(0)]_\rW$. Although the Liouvillian has been truncated in powers of $\hbar$, in general this does not mean that the time-evolved observable has been truncated in $\hbar$, since the action of $\ddp{}{p}$ in the higher-order terms of $\LMoy$ upon the Wigner transformed obervable `brings down' powers of $\hbar^{-1}$ \cite{hel76a}.

The correlation function in \eqr{lscivr} is known as the linearized semiclassical initial value representation (LSC-IVR) or the classical Wigner model, since it can be derived be linearizing the difference in the action between forward-backward trajectories in the semiclassical initial value representation\cite{wan98a}, and was later shown to be derivable from linearizing the action of the exact quantum path-integral\cite{shi03a}. The method is exact in the high-temperature limit, for harmonic systems (where the higher terms in the Moyal series vanish without approximation) and as $t\to 0$\cite{hel15a,shi03a,wan98a}. LSC-IVR gives fairly good short-time dynamics, though can miss interference effects in non-dissipative systems\cite{sun98b,liu15a}. A more serious shortcoming is that the classical dynamics does not conserve the quantum Boltzmann distribution, leading to zero-point energy flowing from high-frequency modes to translations and giving spurious effects in simulations\cite{hab09a}; an effect 
sometimes called `zero-point energy leakage'. Evaluating the Wigner-transformed Boltzmann distribution requires a multidimensional Fourier transform which is often approximated \cite{liu15a}, and at low temperatures this distribution can have negative values \cite{liu09a}. Nevertheless, it has successfully been applied to reaction rates \cite{mar02a}, vibrational energy relaxation and spectra\cite{liu15a,hab09a}.

\section{Matsubara dynamics}
\label{sec:mat}
We have seen how to derive the exact quantum Liouvillian, the Moyal series, and how its truncation to $\mathcal{O}(\hbar^0)$ leads to classical trajectories, though does not conserve the distribution. This motivates considering whether there are other truncations which give classical trajectories (single derivatives in the Liouvillian) but which also conserve the quantum Boltzmann distribution. Here we show that by truncating in the higher path-integral normal modes a classical, Boltzmann preserving `Matsubara' dynamics is produced. Unfortunately it suffers from the sign problem so is not at present a practical method, though we shall subsequently show how its further approximation leads to the successful approximate methods of CMD, RPMD and TRPMD. 

The full derivation of Matsubara Dynamics is in Ref.~\cite{hel15a}; here we outline the necessary steps for a one-dimensional system where $\hat A$ and $\hat B$ are only functions of $\bq$; generalization to more general operators being straightforward\cite{hel15a}. We also require $\hat A$ and $\hat B$ to be invariant w.r.t.\ cyclic permutation of the beads $\{q_i\}$, which is immediately satisfied if $\hat A$ and $\hat B$ are linear as in \eqr{lin}, and is also the case for more general nonlinear operators such as the dividing surface in rate theory\cite{hel13a}. In order to use symmetry w.r.t.\ imaginary time translation, we use the Generalized Kubo Form in \eqr{genk}, insert identities and construct a multidimensional Wigner transform as in \eqr{cwig}, giving\cite{hel15a}
\begin{align}
 \CABNt = \tphN \int d\bq \int d\bp \ [\eb \hat A]_{\bar N}(\bp,\bq) [\hat B(t)]_N(\bp,\bq) \eql{cabgk}
\end{align}
where $N$ is the number of path-integral beads.
The Wigner-transformed Boltzmann distribution is given by 
\begin{align}
 [\eb \hat A]_{\bar N}(\bp,\bq) = & \int d\Delta \ A(\bq) \piNz e^{ip_i\Delta_i/\hbar} \no\\
 & \times  \bra{q_{i-1} - \Delta_{i-1}/2} \ebN \ket{q_i + \Delta_i/2} \eql{anwig}
\end{align}
where the bar on $[\eb \hat A]_{\bar N}$ denotes that the bra-kets link together adjacent [$(i-1)$th and $i$th] beads and the real-time evolution is 
\begin{align}
 [\hat B(t)]_N(\bp,\bq) = & \int d\Delta \int d\bz \ B(\bz) \piNz e^{ip_i\Delta_i/\hbar} \no \\
 & \times \bra{q_i - \Delta_i/2} \etb \kb{z_i}  \etf \ket{q_i + \Delta_i/2} \eql{bnwig}
\end{align}
where the bra-kets only concern a single bead. As all we have done is insert identities, one could equivalently construct \eqr{cabgk} to have $[\eb \hat A]_{ N}(\bp,\bq)$ and $[\hat B(t)]_{\bar N}(\bp,\bq)$. However, since the time-evolution bra-kets only concern a single bead, the Liouvillian for \eqr{cabgk} is simply the sum of the Liouvillian in \eqr{lmoy} acting on each bead:
\begin{align}
 \dd{}{t} [\hat B(t)]_N(\bp,\bq) = \LNMoy [\hat B(t)]_N(\bp,\bq)
\end{align}
where
\begin{align}
 \LNMoy = \smiNz \frac{p_i}{m}\ddp{}{q_i} - \frac{2}{\hbar} V(q_i) \sin\left(\frac{\hbar}{2} \frac{\ola \partial}{\partial q_i} \frac{\ora \partial}{\partial p_i}\right). \eql{lnmoy}
\end{align}
Truncating \eqr{lnmoy} to $\mathcal{O}(\hbar^2)$ gives LSC-IVR in the same way as truncating $\LMoy$ in \eqr{l0} \cite{hel15a}. 

Formally, one can write the exact correlation function in \eqr{cabgk} as
\begin{align}
 \CABNt = \tphN \int d\bq \int d\bp \ [\eb \hat A]_{\bar N}(\bp,\bq) e^{\LNMoy t} B(\bq) \eql{cnelt}
\end{align}
although this will generally be even harder to solve exactly than \eqr{cwig}. The benefit of `repackaging' the correlation function as in \eqr{cnelt} is to exploit its symmetry properties w.r.t.~imaginary time. For example, $[\eb \hat A]_{\bar N}(\bp,\bq)$ and $[\hat B(t)]_N(\bp,\bq)$ (as well as the Liouvillian in \eqr{lnmoy}) are invariant to cyclic permutation in imaginary time (changing $q_i \to q_{i+1}$), whereas this is not obvious with the conventional Kubo-transformed correlation function in \eqr{cwig}. As we shall see later, invariance to translation in imaginary time has a close relationship to the dynamics conserving the quantum Boltzmann distribution.

Instead of writing the correlation function in terms of individual beads, we now consider writing in terms of path-integral normal modes, transforming $(\bq,\bp,\bDelta) \to (\bQ,\bP,\bD)$ where the normal modes are numbered $-(N-1)/2 \le j \le (N-1)/2$ as detailed in Appendix~\ref{ap:nm}\footnote{Here we consider $N$ and $M$ to be odd for algebraic convenience, even $N$ and $M$ leads to the same result \cite{hel15a}.}. In brief, the normal modes conventionally originate from diagonalizing the ring-polymer Hamiltonian (see \eqr{diag} and Ref.~\cite{cer10a}) but here help in evaluating the complex quantum Boltzmann distribution in $[\eb \hat A]_{\bar N}(\bp,\bq)$ and allow an intuitive understanding of the path integral. The lowest mode $Q_0$ is (in this definition) the \emph{centroid} \cite{gil87a,gil87b,vot89a}, the average position of the beads, and $P_0$ the associated momentum. Qualitatively, the modes $Q_{\pm 1}$ describe the size or stretch of the ring polymer \cite{ric09a}, $Q_{\pm 2}$ its curvature and so on. $Q_0$ can therefore be considered the most `classical' of the modes and the modes are more `quantum' with increasing $|j|$.

In normal modes the correlation function becomes \cite{hel15a}
\begin{align}
 \CABNt = \left(\frac{N}{2\pi\hbar}\right)^N \int d\bQ \int d\bP \ [\eb \hat A]_{\bar N}(\bP,\bQ) e^{\LNMoy t} B(\bQ)
\end{align}
where the Liouvillian in normal modes is 
\begin{align}
 \LNMoy = & \smjNN \frac{P_j}{m}\ddp{}{Q_j} \no\\
 & - \frac{2N}{\hbar} \UNQ \sin\left( \frac{\hbar}{2N} \smjNN \frac{\ola \partial}{\partial Q_j} \frac{\ora \partial}{\partial P_j}\right) \eql{lnmoynm}
\end{align}
and the potential in normal modes is given by
\begin{align}
 \UNQ = \frac{1}{N} \smiNz V\!\left( \smjNN \sqrt{N} T_{ij} Q_j \right). \eql{matv}
\end{align}

If we were to truncate \eql{lnmoynm} to $\mathcal{O}(\hbar^0)$ we would recover LSC-IVR once again \cite{hel15a}. Instead, we make a different approximation, truncating from all $N$ to the lowest $M$ path-integral normal modes. From an intuitive perspective, at zero time the highest $N-M$ modes cannot contribute to the (static) correlation function as they are constrained to zero by the quantum Boltzmann operator. One would expect them only to affect the dynamics at longer times when they couple due to anharmonicity in the potential (in a perfectly harmonic potential, the dynamics is separable and the ring polymer normal modes move independently). In the \largeN\ limit, this truncation gives
\begin{align}
 \mL^{[M]} = & \smjMM \frac{P_j}{m}\ddp{}{Q_j} \no\\
 & - \frac{2N}{\hbar} \UNQ \sin\left( \frac{\hbar}{2N} \smjMM \frac{\ola \partial}{\partial Q_j} \frac{\ora \partial}{\partial P_j}\right) \eql{lm}
\end{align}
and we can therefore define an error Liouvillian \cite{hel15b}
\begin{align}
\mL_{\rm er} = \LNMoy - \mL^{[M]} \eql{lerdef} 
\end{align}
which is given in full in appendix~\ref{ap:mater}. 

How many of the lowest $M$ modes should be included? For any physical, analytic potential (one which is smooth, continuous and continuously differentiable) there will be a maximum frequency (second derivative), and provided the frequency of the highest Matsubara mode (see below) is greater than this, all statistical information will be correctly captured (as modes $j \gg M/2$ will move adiabatically to the potential).

For any $M$, the limit $N/M \to \infty$ is taken, and all higher derivatives in \eqr{lm} vanish without approximation, since the $l$th derivative scales as $(M/N)^{l-1}$. Consequently\footnote{Strictly speaking, the potential in \eqr{lm} is $\UNQ$ and this becomes $\UMQ$ after integrating out the non-Matsubara modes detailed below.}
\begin{align}
 \mL^{[M]} = \smjMM \frac{P_j}{m}\ddp{}{Q_j} - \UMQ \smjMM \frac{\ola \partial}{\partial Q_j} \frac{\ora \partial}{\partial P_j} \eql{lm}
\end{align}
and the single derivatives mean that the dynamics is \emph{classical}, with a smoothed ``Matsubara potential'' $\UMQ$.\cite{hel15a}

Because the higher normal modes are not present in the dynamics, nor in $B(\bQ)$, the higher path-integral momenta can be integrated out from the distribution\footnote{This also assumes that $\hat B$ is not a function of the higher normal modes in momenta.}. This allows the higher-frequency `stretch' variables $\{ D_j, |j| > (M-1)/2 \}$ to be integrated out from the distribution. In the \largeN\ limit the Boltzmann bra-kets can be evaluated analytically, leading to the remaining $M$ $\bD$ variables being integrated out by steepest descent. Finally, the higher normal modes in $\bQ$ (which are not affected by $\mL^{[M]}$) can be removed by steepest descent. This leads to the classical-like Matsubara correlation function\cite{hel15a}
\begin{align}
 \CABMt = & \frac{\alpha_M}{2\pi\hbar} \int d\bP \int d\bQ \no\\
 & \times e^{-\beta [H_M(\bP, \bQ) - i \theta_M(\bP, \bQ)]} A(\bQ) e^{\mL^{[M]} t} B(\bQ) \eql{cmat}
\end{align}
where the Matsubara Hamiltonian is 
\begin{align}
 H_M(\bP, \bQ) = \smjMM \frac{P_j^2}{2M} + \UMQ.
\end{align}
The phase factor is given by
\begin{align}
 \theta_M(\bP, \bQ) = \smjMM P_j \tilde \omega_j Q_{-j}
\end{align}
where 
\begin{align}
 \ti \omega_j = \frac{2\pi j}{\beta\hbar}
\end{align}
are the \emph{Matsubara} frequencies\cite{mat55a}, after which the dynamics is named\cite{hel15a}. Note that, in this definition, the frequencies can be negative since $\ti \omega_{-j} = - \ti\omega_{j}$. $\alpha = \hbar^{1-M} [(M-1)/2]!^2$, and the integrals are now implicitly $M$-dimensional as the $N-M$ non-Matsubara modes have been integrated out. 

The truncation in normal modes is illustrated pictorially in \figr{matpic} and mathematically in \figr{matmodes}.
\begin{figure}[h]
\centering
\includegraphics[width=.5\columnwidth]{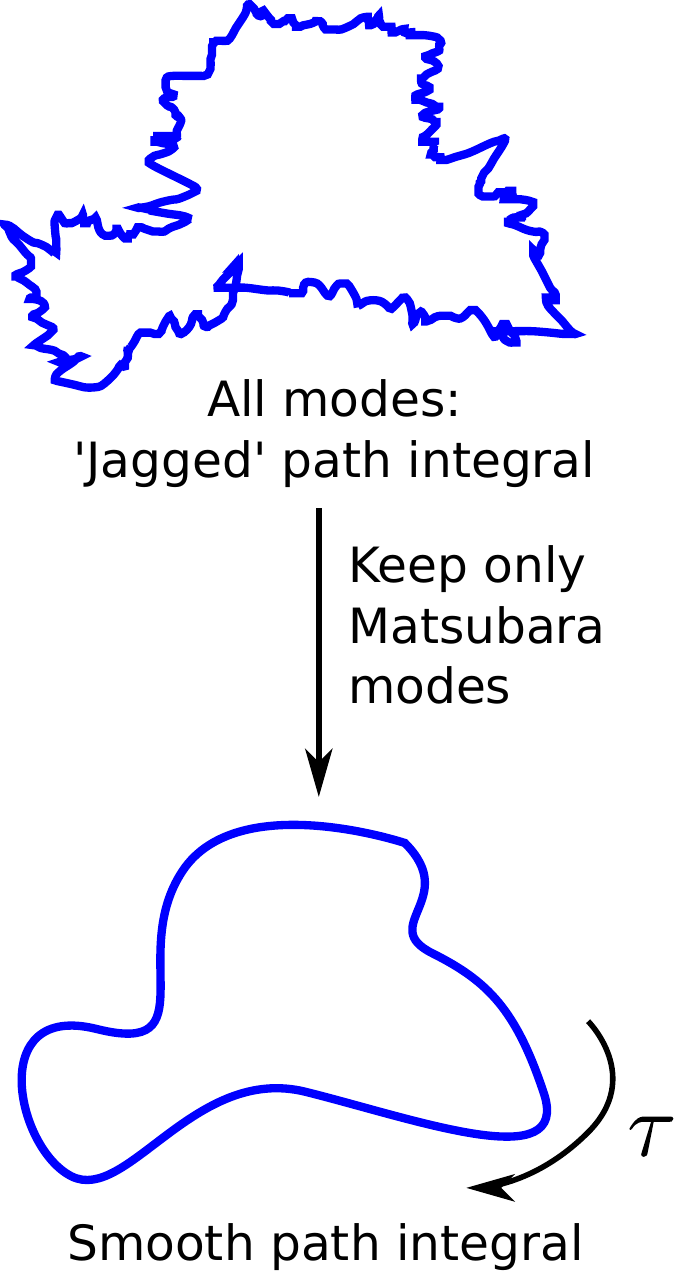}
\caption{Top: inclusion of all path-integral modes leads to a jagged path in imaginary time. Keeping only the lowest Matsubara modes (bottom) leads to a smooth path $\bQ(\tau)$ in imaginary time $\tau$.}
\figl{matpic}
\end{figure}
\begin{figure}[h]
\centering
\includegraphics[width=.9\columnwidth]{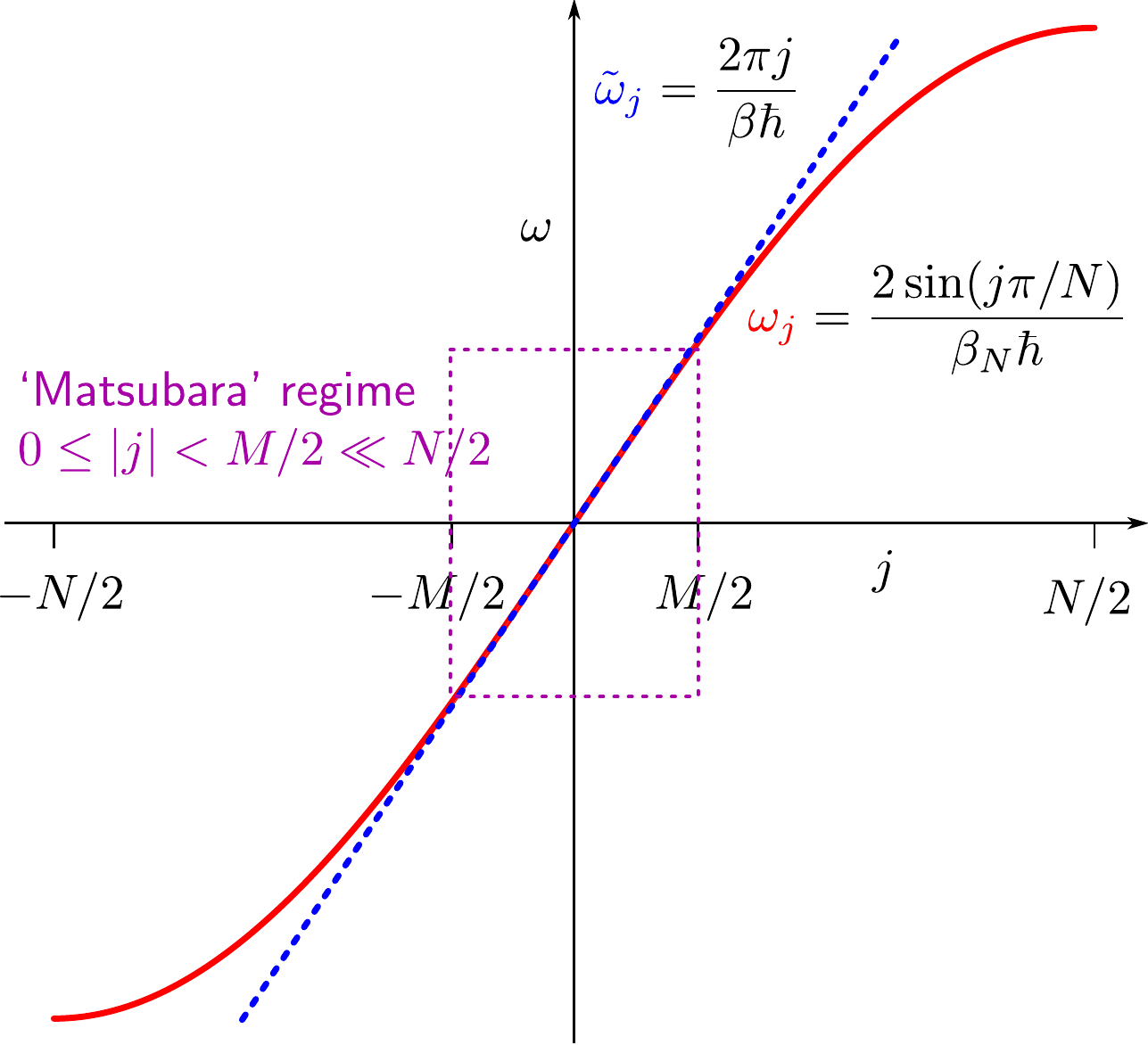}
\caption{Illustrating how the Matsubara modes $\ti\omega_j$ approximate ring polymer modes $\omega_j$ provided $M\ll N$.}
\figl{matmodes}
\end{figure}

Since the dynamics in $\mL^{[M]}$ is equal to that generated by $H_M(\bP, \bQ)$, i.e. $\LM = \{\cdot,H_M\}$ where $\{\cdot,\cdot\}$ is the Poisson bracket, the dynamics will conserve $H_M(\bP,\bQ)$. To show conservation of the phase factor one can either evaluate $\LM \theta_M(\bP,\bQ)$ and show by trigonometric identities that this vanishes, or use Noether's theorem\cite{hel15a}. Using the latter method here, we note that the Hamiltonian and therefore the Lagrangian 
\begin{align} 
\Lambda_M(\bP,\bQ) = \smjMM \frac{P_j^2}{2M} - \UMQ 
\end{align}
is invariant w.r.t.\ translation in imaginary time. Using straightforward differentiation and that $\dd{}{t} Q_j = P_j/m$, 
\begin{align}
 \dd{}{\tau} \Lambda_M = \dd{}{t} \smjMM \left(P_j \dd{Q_j}{\tau}\right) = 0
\end{align}
and by expanding $\dd{Q_j}{\tau}$ in bead co-ordinates and applying trigonometric identities we find 
\begin{align}
 \smjMM \left(P_j \dd{Q_j}{\tau}\right) = \smjMM P_j \tilde \omega_j Q_{-j}
\end{align}
meaning that 
\begin{align}
 \dd{}{t} \theta_M(\bP, \bQ) = 0
\end{align}
and therefore $\LM e^{-\beta [H_M(\bP, \bQ) - i \theta_M(\bP, \bQ)]} = 0$, such that the Matsubara distribution is conserved by the Matsubara Liouvillian, and $\CABMt$ obeys detailed balance.

Matsubara dynamics is therefore classical and conserves the distribution, but the phase factor in the distribution means that the correlation function is not amenable to computation in large systems. However, for the model systems for which it has been computed, it is more accurate than LSC-IVR, CMD or RPMD\cite{hel15a,hel15b}, and is exact for the position-squared correlation function in a harmonic potential\cite{hel16a} which is not the case for RPMD or CMD\cite{hor05a}.

\section{Approximations to Matsubara Dynamics}
\label{sec:approx}
The accuracy of Matsubara dynamics and its intractable nature in large systems suggests that approximations to it which avoid the sign problem may prove more useful in practical applications. Obviously these approximate methods will not in general be as accurate as Matsubara dynamics and one must therefore choose the approximation carefully, in order to remove the sign problem but also keep the dynamics real and preserve the quantum Boltzmann distribution. 

In this article we explore three approximations to Matsubara dynamics which fulfil these criteria; a mean-field approximation which yields centroid molecular dynamics (CMD), and moving the momentum contour in the complex distribution of \eqr{cmat}, followed by approximating the resulting complex dynamics deterministically, giving RPMD, or stochastically, giving TRPMD. The full mathematics is given in a series of recent articles \cite{hel15b,hel16a} and for simplicity only the main details are given here. 

\subsection{Contour integration}
\label{ssec:ci}
For $t=0$, one can perform contour integration in the complex distribution in \eqr{cmat}, defining
\begin{align}
 \bar P_j = P_j - i m \ti\omega_j Q_{-j}
\end{align}
for all the normal modes. There is no phase factor associated with the centroid ($\ti \omega_0 = 0$), and so the countour of the centroid remains unchanged, which will become important later. Using this transformation, for which the Jacobian is unity, we obtain
\begin{align}
 \CABMt = & \frac{\alpha_M}{2\pi\hbar}  \int d\bQ \left[ \pjMM \int_{-\infty - im\ti \omega_jQ_{-j}}^{+\infty - im\ti \omega_jQ_{-j}} d \bar P_j \right] \no \\
 & \times e^{-\beta R_M(\ti \bP, \bQ)} A(\bQ) e^{\mL^{[M]} t} B(\bQ) \eql{contmov}
\end{align}
where $R_M(\ti \bP, \bQ)$ is the ring polymer Hamiltonian in Matsubara modes \cite{hel15b},
\begin{align}
R_M(\ti \bP, \bQ) = & \left(\smjMM \frac{\tilde P_j^2}{2m} + \frac{1}{2}m \ti\omega_j^2 Q_j^2\right) + \UMQ.
\end{align}
In itself, \eqr{contmov} is an exact rewriting of \eqr{cmat}, where $\ti \bP$ are presently complex. However, at zero time, we can evaluate $\{\ti P_j \}$ integrals along the real axis, noting that the edges of the contour vanish, giving
\begin{align}
 C_{AB}^{[M]}(0)= \frac{\alpha_M}{2\pi\hbar} \int d\bQ \int d\ti \bP e^{-\beta R_M(\ti \bP, \bQ)} A(\bQ) B(\bQ), \eql{cabrp}
\end{align}
The contour integral is illustrated pictorially in \figr{contint}.

At finite time, moving the contour in $\{\ti P_j \}$ leads to $\mL^{[M]}$ generating complex trajectories which are inherently unstable \cite{aar10a,aar10c,aar11a,ben08a}, i.e.\ we will have exchanged a complex distribution and real dynamics for a real distribution and complex dynamics, and the problem will be equally (if not more) intractable. However, we will see below that moving the contour and discarding (or replacing) undesirable parts of $\mL^{[M]}$ can lead to tractable dynamics.

\begin{figure}[tb]
\centering
\includegraphics[width=\columnwidth]{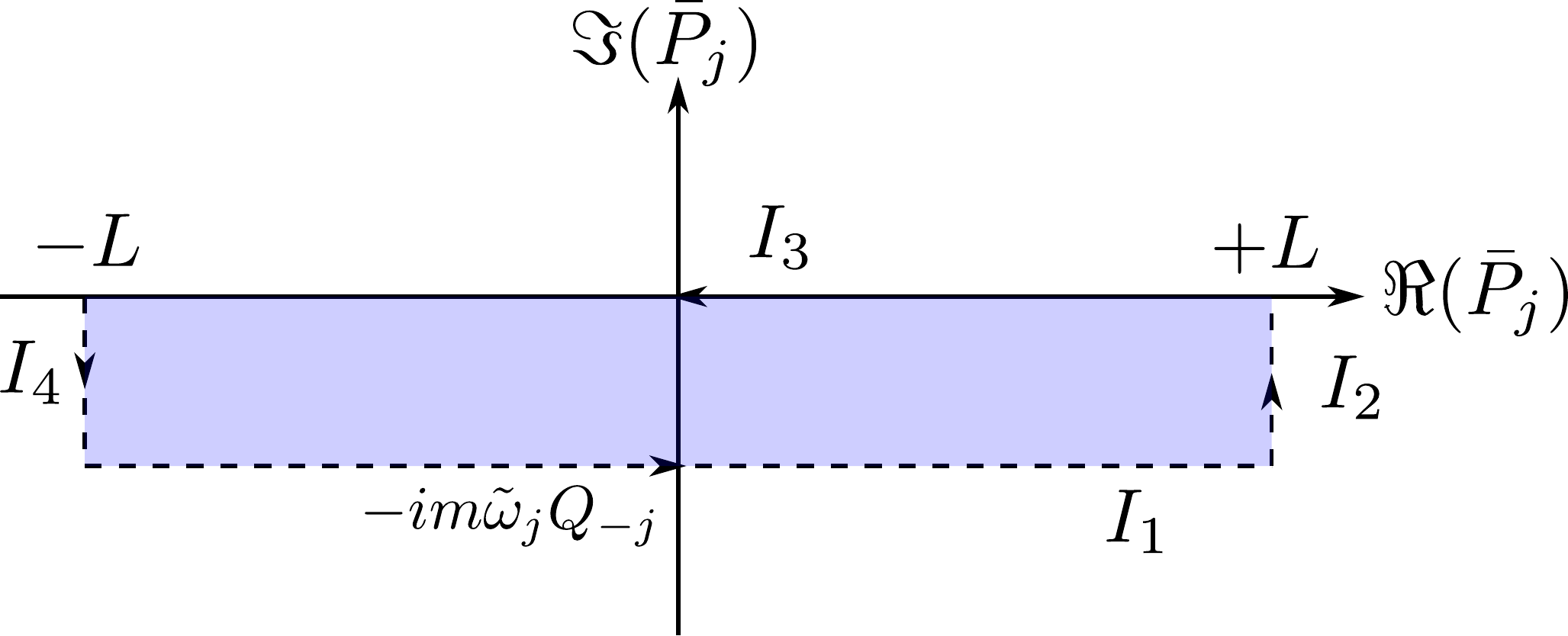}
\caption{The contour integral in \eqr{contmov} in the real and imaginary plane of $\bar P_j$, in the limit $L\to\infty$. Matsubara dynamics evaluates the contour $I_1$, which is real in $P_j$ but complex in $\bar P_j$, and is the contour given in \eqr{contmov}. This is equivalent to $-I_3$, the integral in \eqr{cabrp}, provided that $I_2$ and $I_4$ vanish and that the region enclosed by the contours (shaded blue box, colour online) in holomorphic (free from singularities). This can easily be shown to hold at $t=0$, giving \eqr{cabrp}. At finite time the shaded region is holomorphic for any analytic Hamiltonian \cite{hel16a}, and $I_2$ and $I_4$ can be shown to be zero in a large variety of limits \cite{hel16a}, but their evaluation for an arbitrary potential is challenging \cite{aar10c} and here they are assumed to be zero.}
\figl{contint}
\end{figure}

\subsection{CMD}
\label{ssec:cmd}
If the observables $A(\bQ)$ and $B(\bQ)$ are only functions of the centroid $Q_0$, we formally rewrite \eqr{cmat} as
\begin{align}
  & \CABMt =  \frac{\alpha_M}{2\pi\hbar} \int d P_0 \int d Q_0 \ A(Q_0) \no\\
  & \times \int d\bP' \int d\bQ' \ e^{-\beta [H_M(\bP, \bQ) - i \theta_M(\bP, \bQ)]} e^{\mL^{[M]} t} B(Q_0)
\end{align}
where the primes denote integration over all modes except $P_0$ and $Q_0$. We can then define the reduced centroid density
\begin{align}
 b(Q_0,P_0,t) = & \int d\bP' \int d\bQ' e^{-\beta [H_M(\bP, \bQ) - i \theta_M(\bP, \bQ)]} \no\\
 & \times e^{\mL^{[M]} t} B(Q_0)
\end{align}
and differentiation, followed by integration by parts gives
\begin{align}
 \dd{}{t} b(Q_0,P_0,t) = & \int d\bP' \int d\bQ' e^{-\beta [H_M(\bP, \bQ) - i \theta_M(\bP, \bQ)]} \no\\
 & \times \mL_0 e^{\mL^{[M]} t} B(Q_0) \eql{ddtb}
\end{align}
where the centroid motion alone is given by
\begin{align}
 \mL_0 = \frac{P_0}{m} \ddp{}{Q_0} - \ddp{\UMQ}{Q_0}\ddp{}{P_0} \eql{ml0}
\end{align}
and we have noted that $(\mL^{[M]} - \mL_0)\theta_M(\bP, \bQ) = 0$. At present no approximation has been made and in general direct evaluation of \eqr{ddtb} would be just as difficult as \eqr{cmat} as the force on the centroid in \eqr{ml0} requires evaluting the dynamics of all the other normal modes. However, we can define a mean-field force by averaging over all the non-centroid normal modes,
\begin{align}
 F_0(Q_0) = \frac{-1}{Z_0} \int d\bP' \int d\bQ' e^{-\beta [H_M(\bP, \bQ) - i \theta_M(\bP, \bQ)]} \ddp{\UMQ}{Q_0}
\end{align}
and then perform contour integration as in \eqr{contmov} to obtain
\begin{align}
 F_0(Q_0) = \frac{-1}{Z_0} \int d\bP' \int d\bQ' e^{-\beta R_M(\bP, \bQ)} \ddp{\UMQ}{Q_0}
\end{align}
where the normalization is
\begin{align}
 Z_0 = \int d\bP' \int d\bQ' e^{-\beta R_M(\bP, \bQ)}. \eql{z0}
\end{align}
We can then approximate the force on the centroid as
\begin{align}
 \ddp{\UMQ}{Q_0} = F_0(Q_0) + F_{\rm f} (Q_0) \eql{cenap}
\end{align}
where $F_{\rm f} (Q_0)$ is defined by \eqr{cenap}, and by discarding $F_{\rm f} (Q_0)$ we obtain
\begin{align}
 \dd{}{t} b(Q_0,P_0,t) \simeq  \left[\frac{P_0}{m} \ddp{}{Q_0} + F_0(Q_0) \ddp{}{P_0} \right] b(Q_0,P_0,t)
\end{align}
from which we can define a centroid-only Liouvillian
\begin{align}
\mL_{\rm C} = \frac{P_0}{m} \ddp{}{Q_0} + F_0(Q_0) \ddp{}{P_0}.
\end{align}
and a formal solution 
\begin{align}
b(Q_0,P_0,t) = e^{\mL_{\rm C} t} b(Q_0,P_0,0).
\end{align}
We can now perform the contour integration inside $b(Q_0,P_0,0)$ giving $b(Q_0,P_0,0) = Z_0 B(Q_0)$ where $Z_0$ is the centroid-density distribution given in \eqr{z0}. Since $\mL_{\rm C} Z_0 = 0$, we can `leave' the distribution at zero time and only propagate $B(Q_0)$, giving an approximate correlation function 
\begin{align}
 \CABMt \simeq & \frac{\alpha_M}{2\pi\hbar} \int d P_0 \int d Q_0 \ A(Q_0) Z_0 e^{\mL_{\rm C} t} B(Q_0) \no \\
 = &C_{AB}^{\rm CMD}(t)
\end{align}
which is CMD\cite{vot96a,hel15b,shi03a,jan99b,cao94a,cao94b,cao94c,cao94d}. Consequently, CMD can be obtained from exact quantum dynamics by discarding the motion of the high-frequency modes to obtain Matsubara dynamics, and then making the mean-field approximation $\ddp{\UMQ}{Q_0} \simeq F_0(Q_0)$, i.e.\ that the fluctuations around the centroid are negligible. In some situations such as high temperatures this is a reasonable approximation, but at low temperatures where the ring polymer is highly delocalised this can lead to the curvature problem \cite{wit09a} where spectra are artificially broadened and red-shifted, and reaction rates for asymmetric systems are overestimated since the higher normal modes form part of the optimal dividing surface \cite{ric09a}. Because the higher normal modes are integrated out in CMD, it is inaccurate even at $t=0$ for nonlinear operators \cite{hor05a,ros14a}, though various techniques to address this have been proposed \cite{rei00a,hor05a}.

Because $\mL_{\rm C} Z_0 = 0$, CMD conserves the distribution function and obeys detailed balance.

In theory, there is no mathematical obligation to take the mean field of all non-centroid modes, and one could average out over a subset, such as the most highly oscillatory ones. While this would include some level of fluctuations, the distribution of the non-centroid modes which were not integrated out would still suffer from the sign problem.

\subsection{RPMD}
As noted in section~\ref{ssec:ci}, analytic continuation of the non-centroid momenta is mathematically possible, and the integrand can be proven to be holomorphic in that region of the complex plane\cite{hel16a}, meaning that there are no singularities to worry about. The complex Liouvillian can be written as its real and imaginary parts, 
\begin{align}
\mL^{[M]} = \mL^{[M]}_{\Re} + i\mL^{[M]}_{\Im}
\end{align}
where
\begin{align}
\mL^{[M]}_{\Re} = & \smjMM \frac{\bar P_j}{m}\ddp{}{Q_j} - \left[ m\ti\omega_j^2 Q_j + \frac{\partial  \UMQ }{\partial Q_j} \right] \frac{\partial}{ \partial \bar P_j} \\
= & \mL^{[M]}_{\rm RP}
\end{align}
is the ring polymer Liouvillian (using Matsubara frequencies) and 
\begin{align}
\mL^{[M]}_{\Im} = \smjMM \ti \omega_j \left( \bar P_j \ddp{}{\bar P_{-j}} - Q_j \ddp{}{Q_{-j}} \right). \eql{lim}
\end{align}
One can show that both $\mL^{[M]}_{\Re}$ and $i\mL^{[M]}_{\Im}$ separately conserve the distribution in \eqr{cabrp}, and so discarding $i\mL^{[M]}_{\Im}$ leads to a correlation function with a real distribution and a real dynamics which conserves it,
\begin{align}
 C_{AB}^{\rm RP}(t) = \frac{\alpha_M}{2\pi\hbar} \int d\bQ \int d\ti \bP e^{-\beta R_M(\bar \bP, \bQ)} A(\bQ) e^{\mL^{[M]}_{\rm RP} t} B(\bQ) \eql{rpmd}
\end{align}
which is RPMD\cite{hel15b,cra04a}. This means that the error in the evolution between exact quantum dynamics and RPMD can be stated in closed form as the error between exact quantum dynamics and Matsubara dynamics [\eqr{ler}], followed by a contour integral and discarding $\mL^{[M]}_{\Im}$ [\eqr{lim}]\footnote{Strictly speaking, one also discards the vertical edges of the integral contour, which are believed to be zero\cite{hel16a}.}.

Since $\mL^{[M]}_{\rm RP} e^{-\beta R_M(\bar \bP, \bQ)} = 0$, RPMD conserves the distribution and $C_{AB}^{\rm RP}(t)$ obeys detailed balance. Strictly speaking, \eqr{rpmd} is RPMD with Matsubara frequencies, but in the $M \to \infty$ and $N/M \to \infty$ limits (implicitly taken here), only the lowest Matsubara modes will participate in the statistics and dynamics, the others being constrained to zero by the spring terms in $R_M(\bar \bP, \bQ)$, and correlation functions employing Matsubara and ring polymer frequencies will converge to the same result \cite{hel15b}.

One unfortunate effect of discarding $\mL^{[M]}_{\Im}$ is that it shifts the frequencies of the non-centroid normal modes; in a harmonic potential $V(q) = \frac{1}{2}m\omega_h^2 q^2$, they become \cite{hab13a} 
\begin{align}
\bar \omega_j = \sqrt{\ti \omega_j^2 + \omega_h^2}.
\end{align}
This leads to the so-called `spurious resonances' problem in spectra, where resonances between ring polymer frequencies and physical frequencies (such as stretching vibrations) lead to spurious extra spectra peaks which are temperature-dependent\cite{ros14a,wit09a,iva10a,shi08a}.

\subsection{TRPMD}
To address the artificial shifting of frequencies upon discarding $i\mL^{[M]}_{\Im}$, we consider replacing it with an operator which will conserve the distribution but also provide the correct oscillation frequency. The standard analysis of a damped harmonic oscillator\cite{nit06a} shows that a friction term will reduce the oscillation frequency, so we consider defining\cite{hel16a}
\begin{align}
\mathcal{A}_{\rm RP}^{[M] \dag} = \mL^{[M]}_{\rm RP} + \mathcal{A}_{\rm wn}^{[M] \dag} \eql{aap}
\end{align}
where $\mathcal{A}_{\rm wn}^{[M] \dag}$ is the adjoint of a white-noise Fokker-Planck operator\cite{zwa01a},
\begin{align}
\mathcal{A}_{\rm wn}^{[M] \dag} = - \bar\bP \cdot \bm{\Gamma} \cdot \nabla_{\bar\bP} + \frac{m}{\beta}\nabla_{\bar\bP} \cdot \bm{\Gamma} \cdot \nabla_{\bar\bP}. \eql{wn}
\end{align}
The first term on the RHS of \eqr{wn} corresponds to the drag cause by the semidefinite friction matrix $\bm{\Gamma}$ (which we assume is diagonal in what follows) and the second term represents the `kicks' imparted to the individual momenta of stochastic trajectories\cite{nit06a}. Inserting \eqr{aap} into the analytically continued correlation function gives
\begin{align}
C_{AB}^{\rm TRP}(t) = \frac{\alpha_M}{2\pi\hbar} \int d\bQ \int d\ti \bP \ e^{-\beta R_M(\bar \bP, \bQ)} A(\bQ) e^{ \mathcal{A}_{\rm RP}^{[M] \dag} t} B(\bQ) \eql{trp}
\end{align}
which is TRPMD\cite{hel16a,ros14a}. Similar to RPMD, the approximation in the dynamics between exact quantum evolution and TRPMD is therefore known, namely $\mL_{\rm er}$ followed by a contour integral and replacing $i\mL^{[M]}_{\Im}$ with $\mathcal{A}_{\rm wn}^{[M] \dag}$.

Using integrating by parts one can obtain the (non-adjoint) of the Fokker-Planck operator in \eqr{wn} as \cite{zwa01a}
\begin{align}
\mathcal{A}_{\rm RP}^{[M]} = -\mL^{[M]}_{\rm RP} + \nabla_{\bar \bP} \cdot \bm{\Gamma} \cdot \bar \bP + \frac{m}{\beta}\nabla_{\bar\bP} \cdot \bm{\Gamma} \cdot \nabla_{\bar\bP}
\end{align}
such that the \eqr{trp} can be rewritten as 
\begin{align}
C_{AB}^{\rm TRP}(t) = \frac{\alpha_M}{2\pi\hbar} \int d\bQ \int d\ti \bP \ e^{-\beta R_M(\bar \bP, \bQ)} A(\bQ) e^{ \mathcal{ \ola A}_{\rm RP}^{[M]} t} B(\bQ).
\end{align}
We can then show that $\mathcal{A}_{\rm RP}^{[M]}e^{-\beta R_M(\bar \bP, \bQ)}=0$ such that the stochastic dynamics of the system conserves the distribution. Showing that the correlation function obeys detailed balance is more complicated (since $\mathcal{A}_{\rm RP}^{[M]}$ contains double derivatives) and this is detailed in Ref.~\cite{hel15c}.

Defining the friction matrix to be $\mathbf{\Gamma}_{jk} = 2 |\ti\omega_j |\delta_{jk}$ leads to the correct oscillation frequency of all ring polymer normal modes in a harmonic potential, and therefore give the correct zero-time value and oscillation frequency for the harmonic position-squared autocorrelation function \cite{hel16a}, which neither RPMD nor CMD can achieve \cite{hor05a,hel16a}. More importantly for spectra, a friction matrix of $\mathbf{\Gamma}_{jk} = \sqrt{2} |\ti\omega_j |\delta_{jk}$ will lead all peaks in the position autocorrelation function for a harmonic oscillator to be at the correct (external) frequency, and therefore provides a unique value of $\mathbf{\Gamma}_{jk}$ for computation of spectra which is between the values previously suggested on the basis of optimal sampling\cite{cer10a,ros14a}.

Although TRPMD improves on both CMD and RPMD for spectra \cite{ros14a}, the friction causes unphysical slowing of reaction rates beneath the crossover temperature \cite{hel15c}.

\subsection{Summary}
The various approximations used to obtain LSC-IVR, CMD, RPMD and TRPMD are illustrated schematically in \figr{flow} and their properties summarized in Table~\ref{tab:prop}. For many systems with mild quantum effects some or all of these methods will produce similar results \cite{per09a}, and all are exact in the high-temperature (classical) limit \cite{ros14a,cra04a,liu15a,cao94c}, the $t\to 0$ limit \cite{bra06a,ros14a,liu15a} and for the position autocorrelation function of a harmonic oscillator \cite{liu15a,ros14a,cao94c,cra04a,hel15b}. Although we have shown that CMD can be obtained directly from Matsubara dynamics as a mean field approximation, it can also be obtained as a mean field approximation to RPMD and TRPMD using the same methodology, as shown for RPMD in Ref.~\cite{hon06a}.

\begin{figure*}
\centering
\includegraphics[width=.9\textwidth]{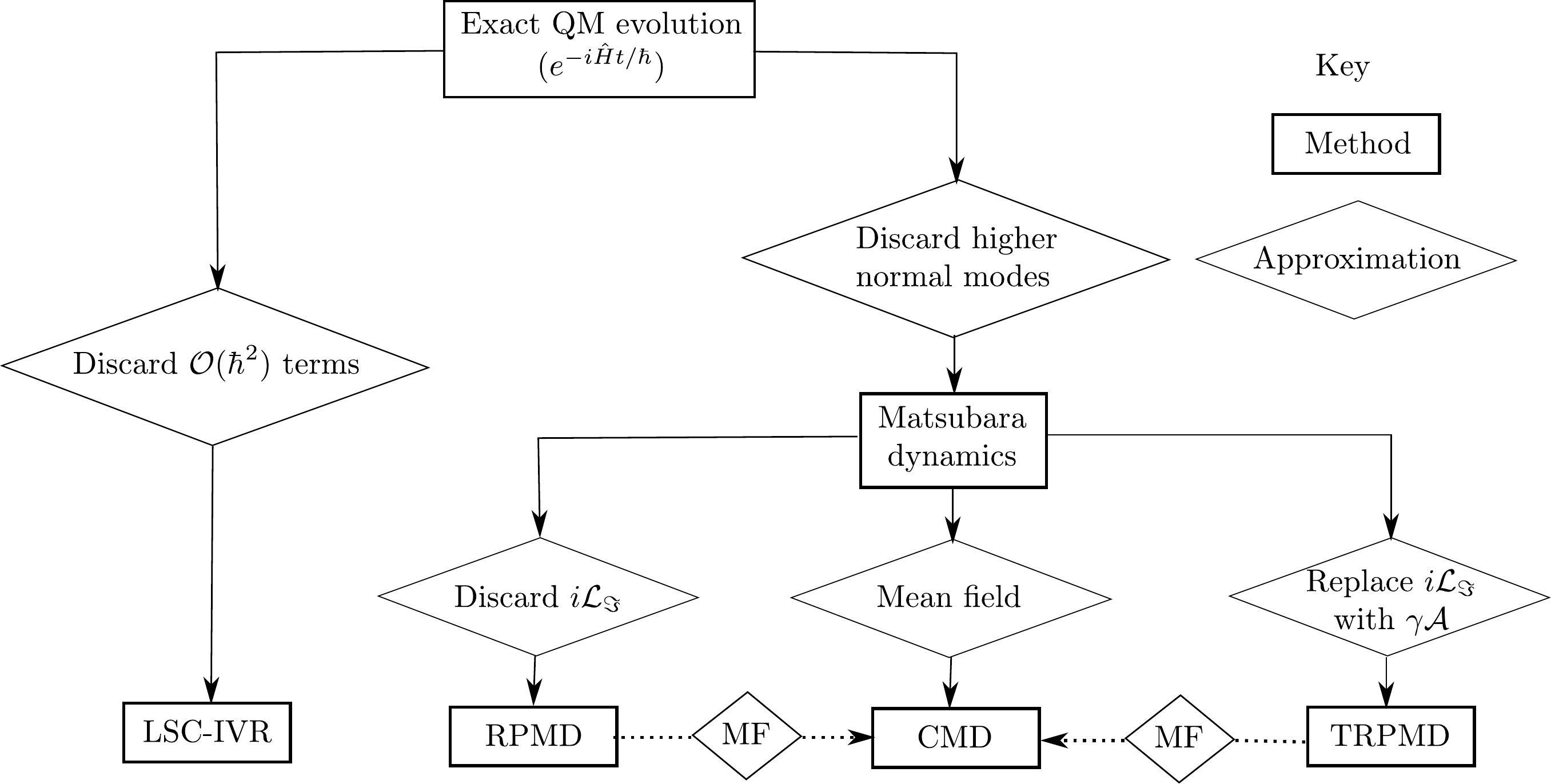}
\caption{Schematic flow diagram illustrating the various approximations from exact evolution to the methods described in the article. MF = Mean field. }
\figl{flow}
\end{figure*}

\begin{table*}
\centering
\begin{tabular}{m{3.5cm}|m{3cm}|m{3cm}|m{3cm}|m{3cm}}
& LSC-IVR & CMD & RPMD & TRPMD \\ \hline 
Approximation & Discard $\mathcal{O}(\hbar^2)$ & Mean field & Discard $i\mL^{[M]}_\Im$ & Replace $i\mL^{[M]}_\Im$ with $\mathcal{A}_{\rm wn}^{[M] \dag}$\\
Conserves distribution and detailed balance? & No & Yes & Yes & Yes \\
Centroid force & N/A & Mean field & Matsubara force & Matsubara force \\
Reaction rates & Problems beneath $T_c$\cite{liu09a} & Inaccurate beneath $T_c$ \cite{jan99a,ric09a} & Good \cite{hab13a,sul16a} & Friction slows rates \cite{hel15c} \\
Spectra & Good\cite{hab09a} & Curvature problem \cite{wit09a,iva10a} & Spurious resonances \cite{wit09a,iva10a} & Good \cite{ros14a} \\
Diffusion & ZPE leakage \cite{hab09a} & Good \cite{ros14a,hab09b} & Good \cite{hab09a} & Good \cite{ros16a} \\
Nonlinear operators & Good if ZPE not problematic \cite{liu15a} & Fails even at $t=0$ \cite{hor05a} & Breakdown from incorrect frequencies \cite{hor05a} & Breakdown from damping \cite{hel16a} \\
Advised usage & Nonlinear operators & Rates above $T_c$, diffusion & Rates, diffusion & Spectra, diffusion
\end{tabular}
\caption{Summary of the properties of LSC-IVR, CMD, RPMD and TRPMD. $T_c$ is the crossover temperature discussed in appendix~\ref{ap:cross}.}
\label{tab:prop} 
\end{table*}

\section{Quantum transition-state theory}
\label{sec:qtst}
Having considered time-correlation functions, we now consider one of their principal applications: reaction rate calculation, and how the foregoing mathematical `toolkit' can be used to obtain quantum transition-state theory.

\subsection{Background}
Here we provide a brief outline of the development of rate theory to place the material discussed here in context; for a fuller historical overview see Ref.~\cite{lai83a}. 

The earliest widely-accepted rate formula is arguably the Arrhenius equation
\begin{align}
k = A e^{-E_a/RT}
\end{align}
where $A$ is the pre-exponential (frequency) factor and $E_a$ is the activation energy. Obtained empirically, there was originally no clear prescription for determining $A$ \emph{a priori}. In 1935 Eyring\cite{eyr35a,eyr35b} along with Evans and Polanyi\cite{eva35a} proposed
\begin{align}
 k=& \sqrt{\frac{m}{2\pi\beta\hbar^2}}K^* \frac{1}{\sqrt{2\pi m \beta}} \kappa \\
 = & \frac{1}{2\pi\beta\hbar}K^* \kappa \eql{tst}
\end{align}
where $\sqrt{\frac{m}{2\pi\beta\hbar^2}} K^*$ is the equilibrium constant between the reactants and the activated complex (the thermal probability of finding the system at the transition state), $1/\sqrt{2\pi m \beta}$ is the thermal flux and, to quote Eyring\cite{eyr35b}
\begin{quotation}
The transmission coefficient $\kappa$ is just the ratio of systems crossing the barrier to systems reacting\ldots
 Fortunately, as stated for many reactions we make a negligible error by taking it as unity.
\end{quotation}
Consequently, \eqr{tst} (hereafter ``Eyring TST'') is the thermal flux multiplied by the probability of forming the activated complex, or in modern terminology, the thermal flux through the dividing surface, which gives the exact rate if there is no recrossing. The partition functions involved are calculated quantum mechanically, but the motion through the transition state is assumed to be classical and separable from motion orthogonal to the dividing surface, which is not always the case \cite{mil76a} and in some circumstances can lead to considerable errors.

\subsection{Classical rate theory}
Determining the functional form of the transmission coefficient was placed on a firmer theoretical footing in the 1970s by constructing a classical flux-side correlation function to determine the classical rate\cite{cha78a,pec73a},
\begin{align}
k_{\rm cl}(\beta)  = \ltti \frac{c_{\rm fs}(t)}{Q_{\rm r}(\beta)} \eql{clrate}
\end{align}
where (in one dimension for simplicity)
\begin{align}
c_{\rm fs}(t) = \tph \int dp \int dq \ e^{-\beta H(p,q)} \delta(q-q^{\ddag}) \frac{p}{m} h(q_t-q^\ddag).\eql{cfs}
\end{align}
This correlates the flux through $\qdd$ at zero time, $\delta(q-q^{\ddag}) p/m$, with whether the system is in the product region at time $t$, $h(q_t-q^\ddag)$. Here $\Qrb$ is the partition function in the reactant region, $\delta(q-\qdd)$ is a Dirac delta function and $h(q_t-q^\ddag)$ a heaviside function, similar to the quantum case. For an $F$-dimensional system one defines a reaction co-ordinate $\fq$ such that $\fq=0$ defines an $(F-1)$-dimensional dividing surface, $\fq>0$ is the product region and $\fq<0$ is the reactant region.

Strictly speaking, the infinite-time limit in \eqr{clrate} is only valid for gas-phase scattering. For condensed-phase systems, in order to define a rate there must be sufficient separation in timescales between reaction and equilibration for plateau in $\cfs$ to emerge, at which point the rate is evaluated\cite{cha78a}. 

\subsection{Classical TST}
Here we show how the classical TST rate is related to the short-time limit of \eqr{cfs} and therefore to the classical rate. In the process we obtain an algebraic expression for the transmission coefficient. We firstly formally rewrite \eqr{cfs} as
\begin{subequations}
\begin{align}
c_{\rm fs}(t) = & \tph \int dp \int dq \ e^{-\beta H(p,q)} \delta(q-q^{\ddag}) \frac{p}{m} e^{\mL t} h(q-q^\ddag) \eql{cfs2} \\
= & \tph \int dp \int dq \ e^{-\beta H(p,q)} \delta(q-q^{\ddag}) \frac{p}{m} h[ (e^{\mL t} q)-q^\ddag] \eql{cfs3}
\end{align}
\end{subequations}
where $\mL$ is the classical Liouvillian given in \eqr{lclas}, and we have used the algebra in Section~\ref{sec:tcf} to take $e^{\mL t}$ `inside' the heaviside function, since $\mL$ only contains single derivatives in $p$ and $q$. Because the heaviside function is discontinuous, one has to be careful expanding $e^{\mL t} h(q-q^\ddag)$ around $t=0$, and it is mathematically simpler to use \eqr{cfs3} rather than \eqr{cfs2}.

In the short-time limit,
\begin{align}
\lttz h[ (e^{\mL t} q)-q^\ddag] = \lttz h[q + pt/m + \mathcal{O}(t^2) - \qdd]
\end{align}
We then note that the Dirac delta function constrains $q = \qdd$ and that the heaviside function is invariant to the scaling of its argument, such that
 \begin{align}
\lttz \delta(q-q^{\ddag})  h[ (e^{\mL t} q)-q^\ddag]  = & \lttz \delta(q-q^{\ddag})  h(pt/m) \no\\ 
= & \delta(q-\qdd) h(p). \eql{shorttc}
 \end{align}
 Putting \eqr{shorttc} back into \eqr{cfs} gives
 \begin{subequations}
  \begin{align}
 \lttz c_{\rm fs}(t) = & \tph \int dp \int dq \ e^{-\beta H(p,q)} \delta(q-q^{\ddag}) \frac{p}{m} h(p) \\
 = & \tph \left[ \int dp\ e^{-\beta p^2/2m} \frac{p}{m} h(p) \right] \no\\
 & \quad \times \left[ \int dq\ e^{-\beta V(q)} \delta(q-q^{\ddag})  \right] \eql{sep}
 \end{align}
 \end{subequations}
 where the integrals in $p$ and $q$ have become separable. The momentum integral is proportional to the thermal flux at inverse temperature $\beta$, and the position integral is proportional to the thermal probability of reaching the transition state $\qdd$. Comparing this with \eqr{tst}, we see that this (suitably scaled by the partition function $\Qrb$) is the classical transition-state theory rate,
\begin{align}
k_{\rm cl}^\ddag(\beta) = \frac{1}{\Qrb} \lttz c_{\rm fs}(t).
\end{align}
The transmission coefficient, which is the ratio of the classical TST rate to the exact classical rate, is therefore given by
\begin{align}
\kappa(t) = \frac{\cfs}{\lim_{t' \to 0_+} c_{\rm fs}(t')} \eql{kappa}
\end{align}
where
\begin{align}
k_{\rm cl}(\beta) = k_{\rm cl}^\ddag(\beta) \times \ltti \kappa(t) \eql{bc}
\end{align}
In practice, rates are often calculated using expressions such as \eqr{bc}, known as the Bennett-Chandler factorization \cite{fre02a}, since this splits the calculation into a statistical part $k_{\rm cl}^\ddag(\beta)$ for which there exists a huge repertoire of efficient sampling techniques\cite{fre02a,tuc10a}, and a dynamical part $\kappa(t)$ which can be obtained from a molecular dynamics simulation. 

From this we can also obtain a mathematical criterion for recrossing. We firstly note that from \eqr{kappa}, $\lttz \kappa(t) = 1$, and obtain the time-derivative of $\kappa(t)$ (c.f.\ \eqr{cff}),
\begin{align}
\dd{}{t} \kappa(t) = \frac{\cff}{\lim_{t' \to 0_+} c_{\rm fs}(t')}
\end{align}
where the classical flux-flux correlation function is 
\begin{align}
\cff = \int dp \int dq \ e^{-\beta H(p,q)} \delta(q-q^{\ddag}) \frac{p}{m} \delta(q_{t}-q^\ddag)\frac{p_{t}}{m}. \eql{cffc}
\end{align}
This gives the flux of particles through the barrier at time $t$, which also went past the barrier at time $t=0$, i.e.\ the extent of recrossing. If there is no recrossing then $\cff = 0$ for all $t > 0_+$, $\kappa(t) = 1$ for all $t\ge 0$, and $k_{\rm cl}(\beta) = k_{\rm cl}^\ddag(\beta)$ which fulfils Eyring's requirement for a TST. 

We can therefore mathematically define classical TST as a rate theory fulfilling two simple criteria:
\begin{enumerate}
\item $k_{\rm cl}^\ddag(\beta) = \frac{1}{\Qrb} \lttz c_{\rm fs}(t)$ such that
\item $k_{\rm cl}^\ddag(\beta) = k_{\rm cl}(\beta)$ if $\cff = 0$ for all $t > 0_+$.
\end{enumerate}
These criteria are not new and are essentially a mathematical summary of the generally-accepted definition of classical transition-state theory \cite{tru80a,cha78a,tru96a,cra05a,fre02a}.

We now briefly note further properties of classical TST which will be useful to compare to QTST. First, if the flux-side time correlation function was defined with two dividing surfaces in different places
\begin{align}
 c_{\rm fs}(t)_2 = \tph \int dp \int dq \ e^{-\beta H(p,q)} \delta(q-q^{\ddag}_1) \frac{p}{m} h(q_t-q^\ddag_2)
\end{align}
where $\qdd_1 \neq \qdd_2$ then
\begin{align}
 \lttz \delta(q-q^{\ddag}_1) h(q_t-q^\ddag_2) = & \lttz \delta(q-q^{\ddag}_1) h(q + pt/m - q^\ddag_2)\no \\
 = & \lttz \delta(q-\qdd) h(q-\qdd_2)
\end{align}
such that
\begin{align}
\lttz c_{\rm fs}(t)_2 = & \tph \int dp \ e^{-\beta p^2/2m} \frac{p}{m} \no \\
& \times \int dq \ e^{-\beta V(q)} \delta(q-\qdd) h(q-\qdd_2) \no \\
= & 0
\end{align}
since the integral in momentum is odd. The existence of a nonzero TST is therefore a consequence of the two dividing surfaces being in the same place \cite{hel13a}.

Second, the separability of the position and momentum terms in the classical TST expression \eqr{sep} means that momentum can be integrated out which (along with evaluating the partition function for a scattering system) gives
\begin{align}
k_{\rm cl}^\ddag(\beta) =  \frac{1}{\sqrt{2\pi \beta m}}\int dq \ e^{-\beta V(q)} \delta(q-\qdd) \eql{noqp}
\end{align}
showing that classical TST does not require the simultaneous specification of position and momentum, even though this is allowed in classical mechanics.

Third, classical rate theory is independent of the location of the dividing surface \cite{mil74a,cra05b}, which can be shown algebraically by differentiating $\cfs$ w.r.t.\ $\qdd$, rearranging, and showing that this corresponds to the system traversing the barrier at time $t$ having starting at the barrier at $t=0$, which cannot be the case at long times if there is a plateau in $\cfs$ and the rate is defined. However, classical TST is exponentially sensitive to the dividing surface. Since recrossing only reduces the rate (by the heaviside function discarding trajectories with positive momentum, or including trajectories with initially negative momentum), classical TST is an upper bound to the classical rate. This property can be used to variationally optimize the location of the dividing surface in multidimensional systems \cite{tru80a}, since in an $F$-dimensional system the dividing surface is an $(F-1)$-dimensional hypersurface, and locating the position of the optimal dividing surface [the one which minimises $k_{\rm cl}^\ddag(\beta)$ and maximises $\kappa(t)$] is difficult.

In summary, classical transition-state theory is the instantaneous thermal classical flux through a position-space dividing surface, which is equal to the exact (classical) rate in the absence of recrossing ($c_{\rm ff}(t > 0) = 0$) by the classical dynamics of the system. It also implicitly assumes that the reactants are in thermal equilibrium (and in equilibrium with the transition state) and that the reaction is electronically adiabatic, proceeding on a single Born-Oppenheimer potential energy surface.\cite{tru96a} The advantages of classical TST over full classical rate calculation is computational simplicity, only requiring knowledge of the PES at the dividing surface and no dynamics, and that it is generally easy to tell in advance if TST will provide a good approximation to the rate. TST works for direct reactions where there is a significant thermal barrier between reactants and products (significantly greater than $k_{\rm B}T$); although it is only exact in a small number of cases (such as one dimensional systems with the optimal dividing surface), recrossing of the optimal dividing surface is often small and it is therefore a good approximation, and upper bound, to the rate.\cite{tru96a,lai83a} It is not expected to work where reactions are diffusive (involving multiple recrossings and therefore a low transmission coefficient), systems with long-lived intermediates (where defining a dividing surface is problematic) or systems with pronounced quantum effects. 

\subsection{Quantum TST}
While very successful for heavy atoms at high temperatures, classical TST does not include any quantum mechanical effects such as tunnelling and zero-point energy, which can lead to significant (many orders of magnitude) deviation between the classical result and the experimental or the quantum result, particularly at low temperatures (see e.g.\ Ref.~\cite{ker06a}). One can, of course, try to include quantum effects into classical TST \cite{tru96a}, such as in the standard Wigner-Eyring model where partition functions in modes orthogonal to the reaction co-ordinate are evaluated quantum mechanically, but motion through the saddle point is assumed to be classical and separable to motion orthogonal to it, which is frequently not the case \cite{mil76a}.

There is considerable historical debate on the existence of quantum transition-state theory, for which the reader is referred to (for example) Refs.~\cite{wig38a,hir39a,mcl74a,mil74a,vot89a,mil93a,vot93a,sma06a}. In short, in the late 1930s Wigner and others considered incorporating quantum effects such as tunnelling into transition-state theory, and noted that there were difficulties due to (a) the non-locality of the quantum Boltzmann operator and (b) the uncertainty principle. 

The non-locality of the quantum Boltzmann operator means that the dividing surface must act on a point or points of the imaginary time trajectory embodied in $\eb$. The development of path-integral techniques by Feynmann \cite{fey65a} and many others means that the dividing surface can be written as a function of path-integral space, $\fq$, taking the positions of path-integral beads $q_1,q_2,\ldots,q_N$ as its argument, such that $\fq=0$ at the dividing surface. To define a rigorous QTST where the only assumption is no recrossing \cite{mil93a} we therefore have to consider recrossing of the path-integral dividing surface $\fq$, and recrossing of any surfaces orthogonal to it in path-integral space, which we denote $\gq$ \cite{alt13a}\footnote{Orthogonality formally means than $\fq \ola \nabla \cdot \ora \nabla \gq = 0$ where $\nabla \gq$ is the gradient of $\gq$ \cite{alt13a}.}.

Concerning the uncertainty principle, specifying the dividing surface in path-integral space allows for a delocalised imaginary-time trajectory and therefore uncertainty in the individual bead positions. We also note that there is no requirement for simultaneous specification of position and momentum in classical TST (see above) and there is no \emph{a priori} reason why this should be required in the quantum case either. 

Extending the definition of classical TST to the quantum case, quantum transition-state theory is therefore defined as the instantaneous thermal flux through a position-dependent dividing surface which gives the exact quantum rate in the absence of recrossing, both of the dividing surface and of the surfaces orthogonal to it in path-integral space\cite{hel13a,alt13a,hel13b,hel14a,hel16b}. Mathematically, we denote $C_{\rm fs}(t)$ to denote a flux-side function correlating flux through $\fq$ at $t=0$ with time-evolved side through $\fq$ (similarly for $C_{\rm ff}(t)$) and $M_{\rm fs}(t)$ for a flux-side function correlating flux through $\fq$ with time-evolved side through $\gq$ \cite{alt13a}. The criteria for QTST given algebraically are therefore
\begin{enumerate}
\item $k^{\ddag}_{\rm Q}(\beta) = \lim_{t\to 0_+} C_{\rm fs}(t)/\Qrb$ such that 
\item $k_{\rm Q}^\ddag(\beta) = k_{\rm Q}(\beta)$ if $C_{\rm ff}(t) = 0$ and $M_{\rm ff}(t) = 0$ for all $t > 0_+$ and all $\gq$.
\end{enumerate}
We stress that the dynamics in these quantum correlation functions is the exact quantum dynamics ($\etf$) and not any of the approximate quantum methods discussed above.

The historical difficulties of formulating a rigorous QTST (satisfying both of the above criteria) led to the development of a huge range of heuristic quantum mechanical rate theories that used transition-state arguments \cite{gil87a,gil87b,vot89a,pol98a,mil74a,wig32b} in addition to alternative approaches such as instanton theory \cite{ric09a,alt11a,zha14b}, quantum instanton methods \cite{mil03a} and many others discussed elsewhere \cite{nym14a,pol05a}. 
There have also been other, generally broader, definitions of QTST in (for example) Refs.~\cite{shi02a,pol05a}. The definition of QTST used in this article is based on Eyring's original definition of TST and means that one has \emph{a priori} knowledge of its applicability: provided there is minimal recrossing QTST will be a good approximation to the rate. 

\subsubsection{Wigner-Miller TST}
Having defined QTST we show how to derive a simple expression satisfying the criteria for a QTST, but which is unreliable at low temperatures. In the followed sections we will extend this to obtain an expression which has positive definite Boltzmann statistics, i.e.\ is guaranteed to be positive at any finite temperature. The original QTST derivation evaluated time-evolution bra-kets algebraically \cite{hel13a}; here we rederive these expressions in the Moyal series formalism, which is arguably simpler. 

As in classical mechanics, the key ingredient in formulating a QTST is ensuring that the two dividing surfaces are located in the \emph{same place} in path-integral space, such that they coalesce in the \shortt\ limit. This has to be done carefully, since the quantum Boltzmann operators is nonlocal, unlike the classical Boltzmann operator. We start with the Wigner-transformed side-side correlation function
\begin{align}
  C_{\rm ss}^{[1]}(t) = &
 \tph \int dq\int dp \ [\eb]_W (p,q) \no \\
 & \times h(q-\qdd) e^{\LMoy t} h(q-\qdd) \eql{ss1}
\end{align}
at $t=0$, the dividing surfaces in \eqr{ss1} are clearly the function of the same co-ordinate and in the same place (they are not separated by an imaginary-time trajectory). 
We obtain the flux-side correlation function as
\begin{align}
C_{\rm fs}^{[1]}(t) = & -\dd{}{t} C_{\rm ss}^{[1]}(t) \no \\
= & \tph \int dq\int dp  \ [\eb]_W (p,q) \no \\
& \times [\LMoy h(q-\qdd)]  e^{\LMoy t} h(q-\qdd) \no \\
= & \tph \int dq\int dp  \ [\eb]_W (p,q) \no\\
& \times \frac{p}{m} \delta(q-\qdd)  e^{\LMoy t} h(q-\qdd) \eql{cfsmoy}
\end{align}
where we have noted that the adjoint of the Liouvillian is its negative\cite{hel16c}, and that $\LMoy [\eb]_W (p,q) = 0$ since exact quantum dynamics conserves the quantum Boltzmann distribution. We illustrate \eqr{cfsmoy} schematically in \figr{qtst1}.

\begin{figure}[tb]
\centering
 \includegraphics[width=.8\columnwidth]{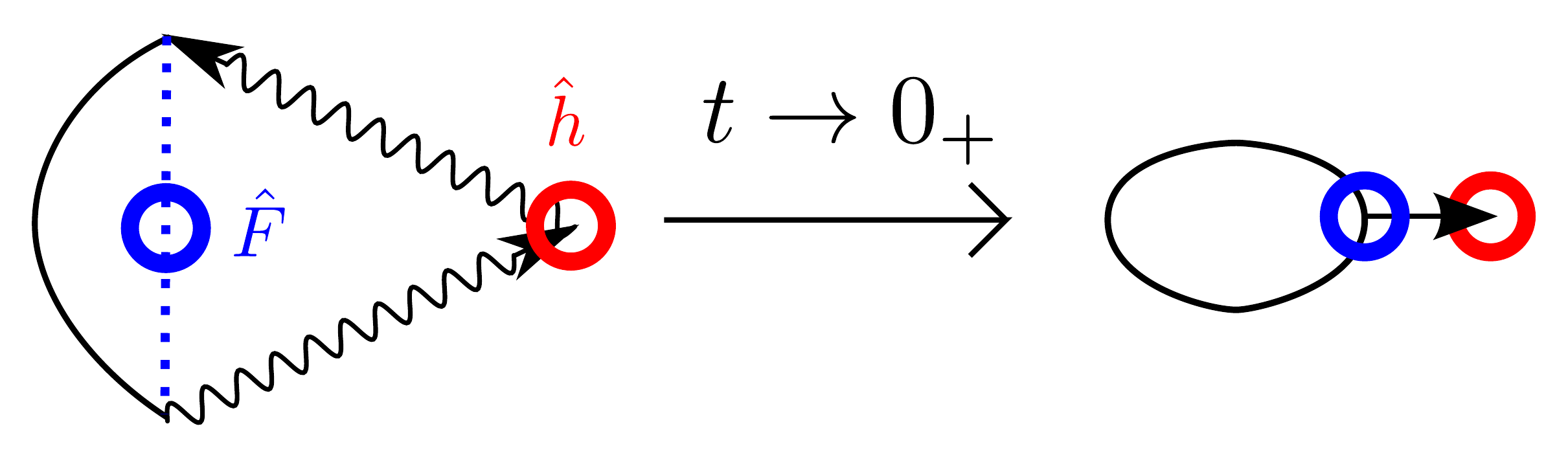}
 \caption{Schematic path-integral diagram of \eqr{cfsmoy} with the imaginary time path shown as a curved line, the real time path with a wavy line and the flux and side operators as blue and red circles respectively. Placing the flux operator at the average of the forward and backward real-time paths (left) leads to the flux and side dividing surfaces being in the same place in path-integral space in the \shortt\ limit (right).}
 \figl{qtst1}
\end{figure}

Expanding $e^{\LMoy t}$ in a Taylor series to find the \shortt\ limit is mathematically problematic since $h(q-\qdd)$ is discontinuous around $q = \qdd$, as for the classical case. However, we can instead write
\begin{align}
 \lttz e^{\LMoy t} = \lttz e^{\mL_{\rm Q}t} e^{\mL_{0}t}
\end{align}
where $\mL_0$ is the classical Liouvillian is defined in \eqr{l0} and $\mL_{\rm Q}$ defined in \eqr{lqdef} contains the higher-order quantum terms. Because $\mL_0$ only contains single derivatives we can use the maths as for the classical case to show
\begin{align}
 \lttz e^{\mL_{\rm Q}t} e^{\mL_{0}t} h(q-\qdd) 
  = & \lttz e^{\mL_{\rm Q}t} h( q + pt/m - \qdd) \no \\
  = & \lttz [1 + \mathcal{O}(t)] h(q + pt/m - \qdd) \no \\
  = & \lttz h(q + pt/m - \qdd)
\end{align}
and therefore
\begin{align}
 \lttz \delta(q-\qdd) e^{\LMoy t} h(q-\qdd) = \delta(q - \qdd) h(p). \eql{delh}
\end{align}
Inserting \eqr{delh} into \eqr{ss1} immediately gives
\begin{align}
C_{\rm fs}^{[1]}(t) = \tph \int dq\int dp  \ [\eb]_W (p,q) \frac{p}{m} \delta(q - \qdd) h(p). \eql{wigrate}
\end{align}
This is identical to a rate expression introduced heuristically by Wigner in 1932 \cite{wig32b} and was subsequently reintroduced and developed for the description of quantum mechanical reaction rates \cite{mil75a,liu09a}. 

The proof that this gives the exact rate in the absence of recrossing is given in \cite{hel13b}, fulfilling the second criterion for a QTST. In brief, since the dividing surface acts only on one point in path-integral space (the average of the end-points of the imaginary time path, see \figr{qtst1}), there are no orthogonal surfaces whose recrossing need be considered. Consequently, as the first criterion for QTST is satisfied, one can combine this with \eqr{cff} to rewrite the second criterion as $\ltti C_{\rm fs}^{[1]}(t)/\Qrb = k_{\rm Q}(\beta)$. This is then proven by evaluating both sides of the equation using quantum scattering theory \cite{hel13b,mil74a,tay06a} where the RHS is given by \eqr{kqb}.

While providing a reasonable description at relatively high temperatures, beneath the `crossover temperature' into deep tunnelling (see appendix~\ref{ap:cross}) the thermal Wigner distribution becomes non-positive definite, such that \eqr{wigrate} can produce spurious negative rates.\cite{liu09a,hel13a} This is because only the \emph{average} of the forward and backward imaginary time paths are constrained to be at the barrier, and the resulting path-integral `string' will sag over the barrier at low temperatures \cite{liu09a,hel13a}.

\subsubsection{Positive-definite statistics}
To ensure that the rate is positive at any finite temperature, the Generalized Kubo correlation function can be used. The full derivation is given in Refs.~\cite{hel13a,alt13a} and here we sketch the pertinent details. A key part of this is defining a dividing surface in path-integral space $\fq$ which must separate the products and reactants, converge with $N$ and (in order to maximise the free energy) a permutationally-invariant function of the path-integral beads \cite{hel13a}. In the terminology of Matsubara dynamics, this means that it must be composed of a finite number of $\mathcal{K}$ Matsubara modes.\cite{hel16b}

We start with the Kubo-transformed side-side correlation function
\begin{align}
\Cssnv{t} = & \tphN \int d\bp \int d\bq\ [\eb]_{\bar N}(\bp, \bq)\no\\
& \times  h[\fq] e^{\LNMoy t} h[\fq]
\end{align}
We then transform the correlation function to path-integral normal modes, without truncating the non-Matsubara modes:
\begin{align}
\Cssnv{t} = & \left(\frac{N}{2\pi\hbar}\right)^N \int d\bP \int d\bQ\ [\eb]_{\bar N}(\bP, \bQ) \no\\
& \times h[\fQ] e^{\LNMoy t} h[\fQ]
\end{align}
As before, we differentiate w.r.t.\ $t$ to obtain the flux-side correlation function
\begin{align}
\Cfsnv{t} = & \left(\frac{N}{2\pi\hbar}\right)^N \int d\bP \int d\bQ\ [\eb]_{\bar N}(\bP, \bQ) \no\\
& \times \delta[\fQ] S(\bP,\bQ) e^{\LNMoy t} h[\fQ] \eql{fsm}
\end{align}
where $S(\bP,\bQ)$ is the ring-polymer flux
\begin{align}
 S(\bP,\bQ) = \frac{1}{m} \sum_{j=-(\mK-1)/2}^{(\mK-1)/2} \ddp{f(\bQ)}{Q_j} P_j 
\end{align}
that is only a function of the lowest $\mK$ normal modes. Equation~\eqref{eq:fsm} and its short-time limit is given schematically in \figr{genkcfs}.

\begin{figure}[tb]
 \centering
 \includegraphics[width=0.5\textwidth]{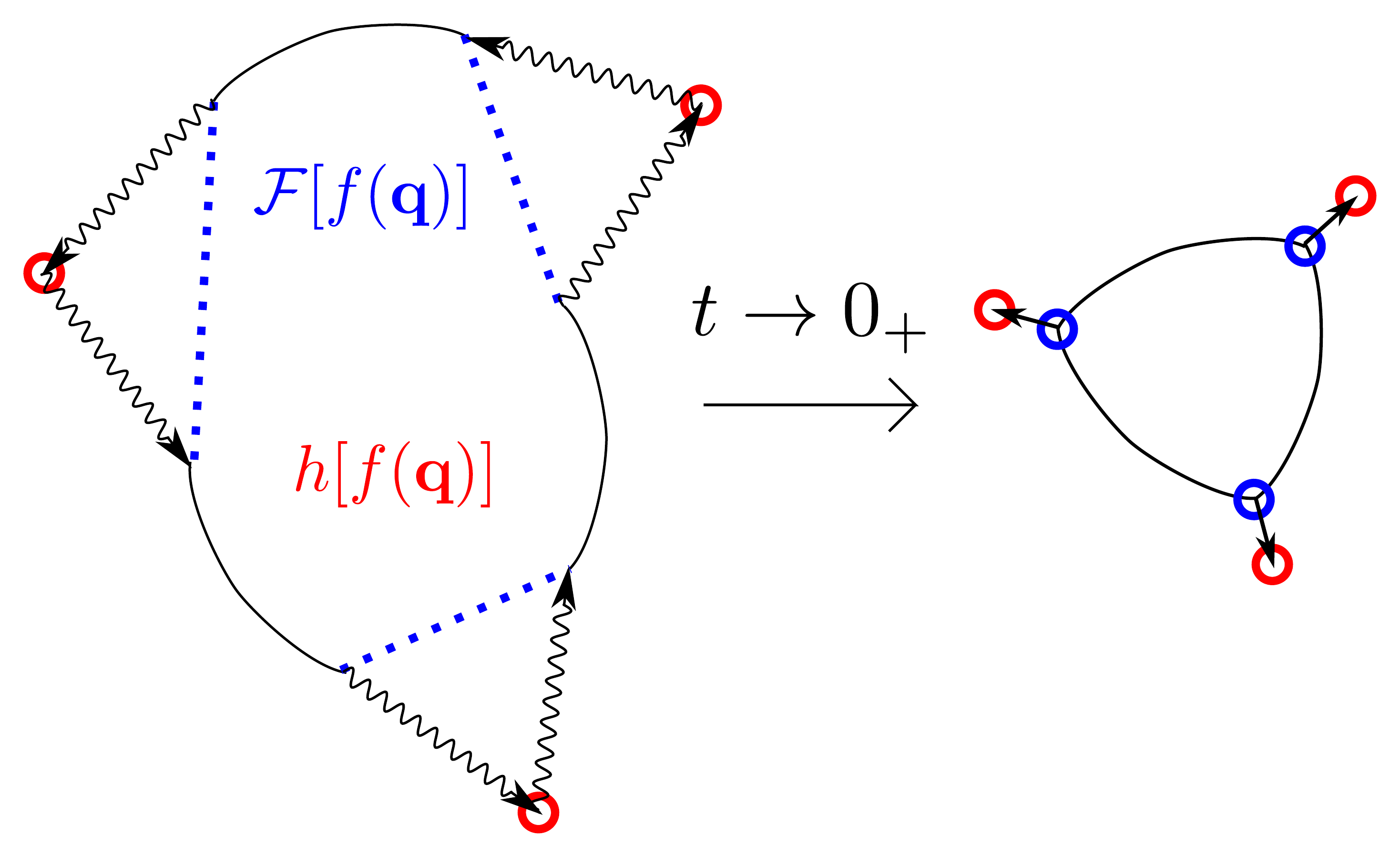}
 \caption{Schematic diagram of the Generalized Kubo correlation function in \eqr{fsm} for the case of $N=3$, with a generalized dividing surface $\fq$ and $\mathcal{F}[\fq] =  \delta[\fQ] S(\bP,\bQ)$ is the flux through $\fq$ at $t=0$. In the \shortt\ limit (and for \largeN), the `stretches' in the ring polymer can be integrated out leading to the ring-polymer flux, shown on the right.}
 \figl{genkcfs}
 \end{figure}
 
In the short-time limit we can separate the propagator
\begin{align}
\lttz e^{\LNMoy t} = e^{\mL_{\rm er}t}e^{\mL^{[M]} t}
\end{align}
where $\mL^{[M]}$ is given in \eqr{lm} and $\mL_{\rm er} = \LNMoy - \mL^{[M]}$, given in appendix~\ref{ap:mater}. For this derivation, we can choose any $M \ge \mK$. Using similar algebra to the classical and Wigner-Miller TST cases, we then show
\begin{align}
\lttz e^{\LNMoy t} h[\fQ] 
= & \lttz h[f(\bQ) + S(\bP,\bQ) t]
\end{align}
where we have Taylor-expanded $f(\bQ)$ and noted that $\mL_{\rm er} h[f(\bQ) + S(\bP,\bQ) t] = 0$ since $h[f(\bQ) + S(\bP,\bQ) t] $ only contains Matsubara modes and all terms in $\mL_{\rm er}$ contain derivatives of non-Matsubara modes. This gives
\begin{align}
\lttz \delta[\fQ] e^{\LNMoy t} h[\fQ] = \delta[\fQ] h[S(\bP,\bQ)] \eql{lttzm}
\end{align}
and inserting \eqr{lttzm} into \eqr{fsm} we obtain
\begin{align}
\lttz \Cfsnv{t} = & \left(\frac{N}{2\pi\hbar}\right)^N \int d\bP \int d\bQ\ [\eb]_{\bar N}(\bP, \bQ) \no\\
& \times \delta[\fQ] S(\bP,\bQ) h[S(\bP,\bQ)], \eql{fqtst}
\end{align}
which is a nonzero \shortt\ quantum transition-state theory by the first criterion, from which we define $k_{\rm Q}^\ddag(\beta) = \lttz \Cfsnv{t}/\Qrb$.

To evaluate \eqr{fqtst} we can, without approximation, integrate out the non-Matsubara $\bP$, followed by $\bD$ inside $[\eb]_{\bar N}(\bP, \bQ)$ and the non-Matsubara $\bQ$ (which by construction are not required to evaluate the distribution) to give 
\begin{align}
\lttz \Cfsnv{t} = & \left(\frac{N}{2\pi\hbar}\right)^N \int' d\bP \int' d\bQ\ e^{-\beta [H_M(\bP, \bQ) - i \theta_M(\bP, \bQ)]} \no\\
& \times  \delta[\fQ] S(\bP,\bQ) h[S(\bP,\bQ)].
\end{align}
This expression is identical to the short-time limit of the Matsubara flux-side time-correlation function, or `Matsubara transition-state theory' (M-TST).  

To address the phase factor, we then move the contour in $\bP$ to generate a ring polymer potential. If the dividing surface contains non-centroid modes we obtain
\begin{align}
S(\bar \bP,\bQ) = \frac{1}{m} \sum_{j=-(\mK-1)/2}^{(\mK-1)/2} \ddp{f(\bQ)}{Q_j} (\bar P_j + i m \ti\omega_j Q_{-j}) \eql{imag}
\end{align}
which appears complex, but the imaginary part corresponds to the change in dividing surface with imaginary time $\tau$, which is zero by construction:
\begin{align}
i \sum_{j=-(\mK-1)/2}^{(\mK-1)/2} \ti\omega_j Q_{-j} \ddp{f(\bQ)}{Q_j}  = & -i \sum_{j=-(\mK-1)/2}^{(\mK-1)/2} \dd{Q_j}{\tau} \ddp{f(\bQ)}{Q_j} \no\\
= & -i \dd{\fQ}{\tau} = 0
\end{align}
where we have used $\ti\omega_j Q_{-j} = - \dd{Q_j}{\tau}$ from Ref.~\cite{hel15a}. This leads immediately to 
\begin{align}
\lttz \Cfsnv{t} = & \left(\frac{N}{2\pi\hbar}\right)^N \int' d\bar \bP \int' d\bQ\ e^{-\beta R_M(\bar \bP, \bQ)} \no\\
& \times \delta[\fQ] S(\bar \bP,\bQ) h[S(\bar \bP,\bQ)] \eql{rptst}
\end{align}
which is RPMD-TST with Matsubara frequencies. As for other static and dynamical properties, this is formally identical to RPMD-TST with ring-polymer frequencies in the large $M$, \largeN\ limit considered here. \cite{hel15a}

We have therefore shown that $\lttz \Cfsnv{t}/\Qrb$ is nonzero giving a QTST by the first criterion. To show that it fulfils the second criterion, we apply \eqr{cff} to the second criterion, and note that $\lttz M_{\rm fs}(t) = 0$ since the dividing surfaces are in different locations in path-integral space. It then becomes sufficient to prove that $\ltti \Cfsnv{t}/\Qrb = k_{\rm Q}(\beta)$ when $\ltti M_{\rm fs}(t) = 0$. The mathematics is given in Ref.~\cite{alt13a}, and in brief the long-time limits are evaluated using quantum scattering theory and we then show that if $\ltti M_{\rm fs}(t) = 0$ for all $\gq$ orthogonal to $\fq$ then $\ltti \Cfsnv{t}$ is equivalent to the long-time limit of $c_{\rm fs}(t)$ in \eqr{cfsq} which by \eqr{kqb} fulfils the second criterion. 

In theory, it is possible to systematically improve QTST to the exact quantum result by computing the recrossing in $\Cfsnv{t}$ and $M_{\rm fs}(t)$ \cite{alt13a,hel14a}, but in practice this is more expensive than a conventional quantum calculation.

\subsubsection{Summary}
We have rederived RPMD-TST and M-TST from a quantum flux-side time-correlation function using the Liouvillian formalism, finding that both are true quantum transition-state theories. Interestingly, for Matsubara TST to be equivalent to QTST only requires that the dividing surface is a function of a finite number of Matsubara modes, but showing the equivalence to RPMD-TST requires the extra condition that the dividing surface is invariant to cyclic permutation.

We also observe that, when the centroid dividing surface is used, RPMD-TST reduces to the earlier centroid-TST \cite{gil87a,gil87b,vot89a,hel13a}. In fact, a recent article claimed to have derived QTST and found that this was equal to Centroid-TST and not RPMD-TST \cite{jan16a}, and which was shown to be an artifact of Ref.~\cite{jan16a} only considering a centroid dividing surface \cite{hel16b}.

In practice, locating the optimal dividing surface $\fq$ is complicated and, particularly at low temperatures, may take on a complicated curvilinear form \cite{ric09a}. Because RPMD rate theory is independent of the location of the dividing surface \cite{cra05b}, the RPMD rate will be equal to the exact quantum rate is there is no recrossing of the optimal dividing surface (the one which minimises $k_{\rm QM}^\ddag$) or those orthogonal to it in path-integral space by either the exact quantum dynamics or the RPMD dynamics of the system. As for classical TST, in general there will be some recrossing, and consequently RPMD is expected to be a good approximation to the rate.

RPMD rate theory itself has seen a huge range of applications, many of which are discussed in Refs.~\cite{hab13a,sul16a}. To mention a few, after initial application to model systems \cite{cra05a,cra05b} it was applied to proton transfer \cite{col08a}, bimolecular reaction rates \cite{col09a,sul11a} and diffusion in ice and clathrates \cite{mar08a,cen16a}. QTST has also been applied to improve standard tunnelling corrections \cite{zha14a}.

Whereas classical TST is an upper bound to the classical rate, QTST is not a strict upper bound to the quantum rate\cite{hel13a}. However, in general QTST is a \emph{good approximation} to an upper bound provided that there are not significant coherences in the reaction dynamics \cite{hel13a}.

\section{Future directions}
\label{sec:fut}

Having surveyed how CMD, RPMD and TRPMD can be considered as approximations to Matsubara dynamics, we briefly consider areas for further development of the field.

\subsection{Nonadiabatic systems}
For small or model systems, exact methods can be applied such as MCTDH \cite{wor08a}, and the past few decades have seen considerably development of approximate methods. 
There exist a wide variety of methods to model non-adiabatic processes using classical-like trajectories, including surface-hopping \cite{tul12a,sub11a,tul90a}, various linearized methods \cite{sun97a}, and mixed quantum-classical \cite{pfa15a,rya14a,wal16a} methods. A common and successful method to map discrete electronic states to continuous classical variables is to use `mapping variables', where singly excited oscillator states are inserted and electronic states represented by their fictitious positions and momenta \cite{mey79a,mey79b,sto97a,sto05a}. There are, of course, many other possible mappings \cite{sto05a} but the simplicity and ease of implementation of mapping variables appears to have led to their widespread application to semiclassical \cite{ana07a,sun98a}, quasiclassical \cite{cot16a}, (partially) linearized \cite{huo11a,huo12a,kim08a,kap06a,hsi12a,hsi13a,bon05a,bon05b}, and path integral dynamics \cite{ric13a,ana13a,duk15a}. Although the propagator (Moyal series) for a single surface systems was obtained in 1949 \cite{moy49a}, the analogue of this in mapping variables was not derived until 2016 \cite{hel16c}.

Despite this progress there remains, to the author's knowledge, no method which has classical-like scaling in all degrees of freedom, conserves the quantum Boltzmann distribution and reproduces Rabi oscillations, though there are a number of methods which incorporate some of these desirable properties\cite{alt16a}. There is also, at present, no widely-accepted `true' (\shortt) non-adiabatic quantum transition-state theory with a dividing surface in electronic space---though this does not mean that one does not exist. For a non-adiabatic system with a dividing surface solely in position space, QTST is simply RPMD-TST with a mean-field non-adiabatic potential \cite{hel14a}, which means that mean-field non-adiabatic RPMD \cite{hel11a,men14a} will provide a good approximation to the exact quantum rate when there is minimal recrossing of the position-space dividing surface by either the (mean field) ring polymer dynamics or the exact quantum dynamics. While this appears to be true for some model systems with large non-adiabatic coupling \cite{hel11a}, this is unlikely to hold in regimes of small coupling \cite{men14a}. Even within existing methods, such as non-adiabatic RPMD, there are a variety of implementations \cite{ric13a,ana13a,duk16a,hel11a,men14a} and it is not always clear which one will be superior in any given situation. 

\subsection{Theoretical development}
There may also be the possibility of applying Matsubara dynamics (or a similar approximate quantum dynamics) to the computation of nonlinear response functions \cite{muk86a} which can diverge in a purely classical calculation \cite{kry06a}. There may also be other classical-like approximations to quantum dynamics (and maybe Matsubara dynamics) that for some systems are more accurate\cite{has16a}. Very recent research has obtained out-of-equilibrium RPMD and CMD from Matsubara dynamics \cite{wel16a}, which should be useful tools for excited state quantum dynamics.
 
\subsection{Computational development}
For a method to bridge the gap between theoretical development and routine application in large chemical systems, the speed of computation needs to be comparable to that of a standard classical molecular dynamics simulation. There have consequently been a large range of methods developed to implement the approximate methods described here accurately and efficiently.

For single-surface systems, there have been impressive applications including a study of dynamics and dissipation in enzyme catalysis \cite{boe11a} and proton transport in water nanowires \cite{ros16a}, though applications to large systems are often limited by the cost of the potential. Various techniques have evolved to address this, including ring polymer contraction \cite{mar08a,mar16a,wan16a} and thermostatting \cite{cer10a,ros14a,hel16a}. 

Open source codes such as i-Pi \cite{cer13b} and RPMDrate \cite{sul13a} have been developed to facilitate application to wide-ranging systems.

\section{Conclusions}
\label{sec:con}
In this New View we have reviewed how a number of successful approximate quantum dynamics methods can be obtained from exact quantum time evolution and used the Liouvillian and Moyal formalisms to rederive quantum transition-state theory. 

We have mainly considered the mathematical basis for these theories and shown what terms they discard from the exact quantum evolution to obtain a classical-like dynamics from which to compute a correlation function. Provided the discarded error terms are small, the approximate correlation function will be a good approximation to the exact quantum correlation function. By considering cases where this is (and is not) the case, we can propose \emph{a priori} situations where a particular methods is likely to work, and therefore advise the usage of approximate methods, summarized in Table~\ref{tab:prop}.

We then revisit classical and quantum transition-state theory and derive QTST in the Matsubara formalism, showing that Matsubara-TST is a true QTST. Provided the dividing surface is permutationally invariant, RPMD-TST is equivalent to Matsubara-TST, unlike the dynamics in RPMD which is only an approximation to Matsubara dynamics. While of limited computational importance by itself (due to the phase factor in the Matsubara distribution) this may facilitate the derivation of other (possibly more accurate) rate theories.

While there has been much progress in recent years, there remain many avenues for further theoretical development. There is arguably no clear consensus on how to apply approximate path-integral methods to non-adiabatic systems, nor a \shortt\ non-adiabatic QTST, the existence of which is an open question. There is also scope for applying the approximate methods discussed here to out-of-equilibrium systems and nonlinear response functions, in addition to developing efficient computational algorithms for implementing these methods in code libraries and for large systems.

\section{Acknowledgements}
TJHH wishes to thank Jesus College, Cambridge for funding, Stuart Althorpe for helpful discussions, and Srinath Ranya and Elliot C.\ Eklund for comments on the manuscript.

\appendix
\section{Ring Polymers}
\label{ap:rp}
There exists a vast literature on ring polymers \cite{fey65a,kle09a} and here we give the standard derivation of the expression for a partition function \cite{tuc10a} for the benefit of those unfamiliar or new to the subject.

For a quantum mechanical partition function
\begin{align}
Z = \tr [\eb]
\end{align}
we can perform the Trotter discretization
\begin{align}
Z = \tr [(\ebN)^N]
\end{align}
and in the \largeN\ limit, expand $\ebN$ symmetrically as
\begin{align}
Z = \lNti \tr [ (e^{-\betaN \hat V/2} e^{-\betaN \hat T} e^{-\betaN \hat V/2} )^N]
\end{align}
where $\hat V = V(\hat q)$ and $\hat T = \hat p^2/2m$. We then insert $N$ sets of position identities, $\int d q_i \kb{q_i} $, $i = 1,\ldots,N$, 
\begin{subequations}
\begin{align}
Z = & \lNti \int d\bq \piN \bra{q_{i-1}} e^{-\betaN \hat V/2} e^{-\betaN \hat T} e^{-\betaN \hat V/2} \ket{q_i} \eql{z1} \\
= & \lNti \int d\bq \piN \bra{q_{i-1}} e^{-\betaN \hat T} \ket{q_i} e^{-\betaN V(q_i)} \eql{z2}
\end{align}
\end{subequations}
where we have noted $e^{-\betaN \hat V/2} \ket{q_i} = \ket{q_i} e^{-\betaN V(q_i)/2}$ and cyclic permutation within indices to go from \eqr{z1} to \eqr{z2}. By inserting momentum eigenstates, we then evaluate
\begin{align}
\bra{q_{i-1}} e^{-\betaN \hat T} \ket{q_i} = & \int dp_i \bk{q_{i-1}}{p_i} e^{-\betaN p_i^2/2m} \bk{p_i}{q_i} \no \\
= & \tph \int dp_i e^{i p_i (q_{i-1} - q_i)/\hbar} e^{-\betaN p_i^2/2m} \no \\
= & \sqrt{\frac{m}{2\pi\betaN\hbar^2}} e^{-m(q_i - q_{i-1})^2/2\betaN\hbar^2} \eql{qTq}
\end{align}
by contour integration, and by inserting \eqr{qTq} into \eqr{z2} obtain
\begin{align}
Z = \lNti \left(\frac{m}{2\pi\betaN\hbar^2}\right)^{N/2} \int d\bq \ e^{-\betaN U_N(\bq)}
\end{align}
where the ring polymer potential is
\begin{align}
U_N(\bq) = \smiN V(q_i)  + \frac{m(q_i - q_{i-1})^2}{2\betaN^2\hbar^2}
\end{align}
One can re-insert $N$ momentum identities\cite{par84a}
\begin{align}
1 = \sqrt{ \frac{\betaN}{2\pi m} } \int dp \ e^{\betaN p^2/2m}
\end{align}
in $p_i$, $i = 1,\ldots,N$ to give
\begin{align}
Z = \lNti \tphN \int d\bq \int d\bp \ e^{\betaN R_N(\bp,\bq)}
\end{align}
where the ring polymer Hamiltonian is
\begin{align}
R_N(\bp,\bq)  = \smiN \frac{p_i^2}{2m} + V(q_i)  + \frac{m(q_i - q_{i-1})^2}{2\betaN^2\hbar^2}. \eql{rpham}
\end{align}
The above derivation is exact for \emph{static} properties and the dynamics generated by \eqr{rpham} was originally proposed as a sampling tool\cite{par84a}. The $\{q_i\}$ are known as ring polymer `beads' and in practice their number $N$ is treated as a convergence parameter in a numerical simulation.

\section{Normal modes}
\label{ap:nm}
The ring-polymer normal modes are defined here as in Ref.~\cite{hel15b},
\begin{align}
 Q_j = \smiNz \frac{T_{ij}}{\sqrt{N}} q_i 
\end{align}
where $j = -N/2+1, \ldots,0,\ldots, N/2$ and likewise for $\bP$, where
\begin{align}
 T_{ij} = 
 \left\{
 \begin{array}{ll}
  N^{-1/2} & j=0 \\
  \sqrt{2/N} \sin(2\pi ij/N) & 1 \leq j \leq N/2 - 1 \\
  N^{-1/2} (-1)^i & j = N/2 \\
  \sqrt{2/N} \cos(2\pi ij/N) & -N/2 + 1 \leq j \leq - 1
 \end{array}
 \right.
\end{align}
where the $j=N/2$ mode is omitted if $N$ is odd. The transformation is not unitary, but defined such that the normal modes converge in the \largeN\ limit. This leads to frequencies in the complex Boltzmann distribution of
\begin{align}
 \omega_j = \frac{2 \sin(j\pi/N)}{\betaN\hbar} \eql{rpfreq}
\end{align}
which, for large $N$ and finite $j$, become the Matsubara frequencies \cite{mat55a}
\begin{align}
 \tilde \omega_j = \lNti \omega_j = \frac{2\pi j}{\beta \hbar}.
\end{align}
The observables $A(\bQ)$ and $B(\bQ)$ are obtained by making by substituting
\begin{align}
 q_i = \smjMM T_{ij} \sqrt{N} Q_{j}
\end{align}
into $A(\bq)$ and $B(\bq)$ respectively, which also leads to a `Matsubara potential' in \eqr{matv}. This transformation also diagonalizes the spring part of the ring polymer Hamiltonian in \eqr{rpham},
\begin{align}
\smiN  \frac{m(q_i - q_{i-1})^2}{2\betaN^2\hbar^2} = \frac{Nm}{2}\omega_j^2 Q_j^2 \eql{diag}.
\end{align}

\section{Matsubara error Liouvillian}
\label{ap:mater}
By exploiting trigonometric identities, \eqr{lerdef} can be given as \cite{hel15a}
\begin{align}
\mL_{\rm er} = & \sum_{j=(M+1)/2}^{(N-1)/2} \frac{P_j}{m}\ddp{}{Q_j} + \frac{P_{-j}}{m}\ddp{}{Q_{-j}} \no\\
& -\frac{4}{\hbar}\UNQ \sin\!\left(\frac{\hat X}{2}\right) \cos\!\left(\frac{\hat X}{2} + \hat Y\right) \eql{ler}
\end{align}
where $\hat X$ acts only on the non-Matsubara modes
\begin{align}
\hat X = \frac{\hbar}{2} \sum_{j=(M+1)/2}^{(N-1)/2} \frac{\ola \partial}{\partial Q_j}\frac{\ora \partial}{\partial P_j} + \frac{\ola \partial}{\partial Q_{-j}}\frac{\ora \partial}{\partial P_{-j}}
\end{align}
and $\hat Y$ acts on the Matsubara modes
\begin{align}
\hat Y = \frac{\hbar}{2}  \smjMM \frac{\ola \partial}{\partial Q_j}\frac{\ora \partial}{\partial P_j}.
\end{align}
Although $\mL_{\rm er}$ contains both Matsubara and non-Matsubara derivatives, expanding the trigonometric functions in \eqr{ler} shows that all terms in $\mL_{\rm er}$ contain at least one derivative in a non-Matsubara mode.

\section{Crossover temperature}
\label{ap:cross}
A rough guide for the temperature beneath which quantum effects become pronounced is the crossover temperature where the first ring polymer normal mode becomes unstable, defined as \cite{ric09a}
\begin{align}
 T_c = \frac{\hbar\omega_b}{2\pi k_{\rm B}}
\end{align}
where $\omega_b$ is the imaginary frequency at the top of the barrier. Since at the maximum $dV(q)/dq = 0$ by construction, the potential can be expanded as $V(q-\qdd) \simeq V(\qdd) - m\omega_b^2 q^2/2 + \mathcal{O}(q^3)$, and $\omega_b$ therefore provides a guide concerning how `peaked' the barrier is, as sketched in \figr{barrier}.
\begin{figure}[h]
\centering
 \includegraphics[width=.2\textwidth]{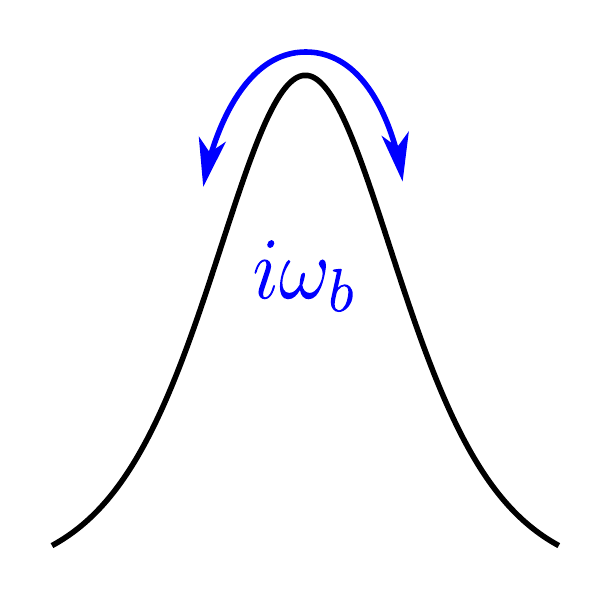}
 \caption{Schematic diagram showing the imaginary mode at the top of the barrier, which would be a saddle point on a multidimensional potential energy surface.}
 \figl{barrier}
\end{figure}

\bibliography{refbig}
\end{document}

%% file: abbrev.tex
\newcommand{\cff}{c_{\rm ff}(t)}

\newcommand{\cfs}{c_{\rm fs}(t)}

\newcommand{\Cssnv}[1]{C_{\rm ss}^{[N]}(#1)}
\newcommand{\Cfsnv}[1]{C_{\rm fs}^{[N]}(#1)}

\newcommand{\CABNt}{C_{AB}^{[N]}(t)}
\newcommand{\CABMt}{C_{AB}^{[M]}(t)}
\newcommand{\eb}{e^{-\beta \hat H}}

\newcommand{\ebN}{e^{-\beta_N \hat H}}

\newcommand{\etf}{e^{-i \hat H t/\hbar}}
\newcommand{\etb}{e^{i \hat H t/\hbar}}

\newcommand{\bra}[1]{\langle #1 |}
\newcommand{\ket}[1]{| #1 \rangle}
\newcommand{\kb}[1]{\ket{#1}\bra{#1}}
\newcommand{\bk}[2]{\bra{#1} #2 \rangle}

\newcommand{\dd}[2]{\frac{d #1}{d #2}}

\newcommand{\ddp}[2]{\frac{\partial #1}{\partial #2}}

\newcommand{\piN}{\prod_{i=1}^N}
\newcommand{\piNz}{\prod_{i=0}^{N-1}}

\newcommand{\smiN}{\sum_{i=1}^N}
\newcommand{\smiNz}{\sum_{i=0}^{N-1}}

\newcommand{\smkNz}{\sum_{k=0}^{N-1}}

\newcommand{\smjMM}{\sum_{j=-(M-1)/2}^{(M-1)/2}}

\newcommand{\smjNN}{\sum_{j=-(N-1)/2}^{(N-1)/2}}

\newcommand{\pjMM}{\prod_{j=-(M-1)/2}^{(M-1)/2}}

\newcommand{\lttz}{\lim_{t\to 0_+}}
\newcommand{\ltti}{\lim_{t\to \infty}}

\newcommand{\lNti}{\lim_{N\to \infty}}
\newcommand{\shortt}{$t\to 0_+$}

\newcommand{\largeN}{$N\to\infty$}

\newcommand{\tr}{ {\rm Tr} }
\newcommand{\qdd}{q^\ddag}

\newcommand{\Qrb}{Q_{\rm r}(\beta)}

\newcommand{\no}{\nonumber}

\newcommand{\betaN}{\beta_N}
\newcommand{\ola}{\overleftarrow}
\newcommand{\ora}{\overrightarrow}

\newcommand{\ti}{\tilde}

\newcommand{\rW}{\mathrm{W}}
\newcommand{\bp}{ {\bf p} }

\newcommand{\bq}{ {\bf q} }

\newcommand{\bz}{ {\bf z} }
\newcommand{\bDelta}{ \bm{\Delta} }

\newcommand{\bQ}{ {\bf Q} }

\newcommand{\bP}{ {\bf P} }
\newcommand{\bD}{ {\bf D} }

\newcommand{\mK}{\mathcal{K}}

\newcommand{\fq}{ f({\bf q}) }
\newcommand{\fQ}{ f({\bf Q}) }

\newcommand{\gq}{ g({\bf q}) }

\newcommand{\tph}{\frac{1}{2\pi\hbar}}

\newcommand{\tphN}{\frac{1}{(2\pi\hbar)^N}}

\newcommand{\inti}{\int_{-\infty}^{\infty}}
\newcommand{\eqr}[1]{Eq.~\eqref{eq:#1}}

\newcommand{\eql}[1]{\label{eq:#1}}
\newcommand{\figr}[1]{Fig.~\ref{fig:#1}}
\newcommand{\figl}[1]{\label{fig:#1}}

\newcommand{\LMoy}{\mathcal{L}_{\rm Moy}}

\newcommand{\LNMoy}{\mathcal{L}_{\rm Moy}^{[N]}}

\newcommand{\mL}{\mathcal{L}}

\newcommand{\UNQ}{U^{[N]}(\bQ)}

\newcommand{\UMQ}{U^{[M]}(\bQ)}